\documentclass[aps,pre,twocolumn,amsmath,amssymb,showpacs,amsfonts]{revtex4}
\usepackage{epsfig}
\usepackage{graphicx}
\usepackage{dcolumn}
\usepackage{bm}

\begin{document}

\title{Inhomogeneous quasi-stationary state of dense fluid of inelastic hard spheres}

\author{Itzhak Fouxon}
\affiliation{Department of Complex Systems,
Weizmann Institute of Science, Rehovot 76100, Israel}
\date{\today}

\begin{abstract}

We study closed {\it dense} collections of hard spheres that collide inelastically with constant coefficient of normal restitution. We find inhomogeneous states (IS) where the density profile is spatially non-uniform but constant in time. 
The states are exact solutions of non-linear partial differential equations that describe the coupled distributions of density and temperature when inelastic losses of energy per collision are small. 
The derivation is performed without modelling the equations' coefficients that are unknown in the dense limit (such as the equation of state),
using only their scaling form specific for hard spheres. The IS is exact non-linear state of this many-body system. It captures a fundamental property of inelastic collections of particles: the possibility of preserving non-uniform temperature via the interplay of inelastic cooling and heat conduction, generalizing previous results in the dilute case. We perform numerical simulations to demonstrate that arbitrary initial state evolves to the IS in the limit of long times where the container has the geometry of the channel. The evolution is like gas-liquid transition. The liquid condenses in a vanishing part of the total volume but takes most of the mass of the system. However, the gaseous phase, which mass grows only logarithmically with the system size, is relevant because its fast particles carry most of the energy of the system. 
Remarkably, the system self-organizes to dissipate no energy: the inelastic decay of energy is a power-law $[1+t/t_c]^{-2}$ where $t_c$ diverges in the thermodynamic limit. This behavior is caused by unusual spatial distribution of particles: on approach to one of the container's walls the density grows inversely with the distance. 
We discuss the relation of our results to the recently proposed finite-time singularity in other container's geometries.

\end{abstract}

\maketitle

\section{Introduction}

In the past decades a lot of attention was devoted to the study of formation of structures in closed systems with dissipative interactions (ordering). This formation is often associated with decrease of entropy due to the the system's interaction with the environment. The latter is described effectively by the dissipative part of the interactions. 
In this work we consider the fundamental dissipative system of hard spheres with inelastic collisions. This serves as basic model of the granular material, which is the dissipative system of macroscopic particles which interaction involves friction, like sand. 
We demonstrate the formation of stable spatial structure in this system. 

The inelasticity of the collisions of the spheres, which model the sand grains, mimics the friction, describing effectively the transfer of energy to the inner degrees of freedom of the particles, that play the role of the environment for the translational degrees of freedom \cite{BP,Goldhirsch2}. 
We use the popular model where the normal component of the relative velocity of the particles is depleted after the collision by a positive constant $r<1$ (which is called coefficient of normal restitution; $r=1$ for the elastic collisions). 

We consider the case where the inelasticity is small, $1-r\ll 1$, so the energy is "almost" conserved in each collision. In this case, the inelasticity effect becomes significant only after long time of evolution. Over the mean free time (or the liquid relaxation time in the dense regions), however, the inelasticity is negligible so the collection reaches the state of the local thermal equilibrium \cite{Landau,Landau10}. That is characterized by the local values of the density, temperature and velocity that vary throughout the system and evolve according to the equations of the fluid mechanics. 

The fluid-mechanical equations of the collection of inelastically colliding hard spheres contain a correction due to inelasticity. That brings finite effect over the time-scale of the order of $(1-r)^{-1}$ times the mean free time. During this time-scale the energy depletion due to the inelasticity is of order one, so the system's state is completely different from the one of the elastic hard spheres.  

In this work we use the fluid mechanical equations of granular materials to demonstrate new states of {\it dense} collections of inelastically colliding hard spheres where the liquid phase is present in the system. These states are states of mechanical equilibrium where the pressure is spatially uniform (this uniform pressure decays in time though due to inelasticity). The forces on each fluid element are balanced so the fluid is macroscopically at rest and the density profile is time-independent. Thus these states can be said to be closest to the equilibrium states of collections of elastically colliding hard spheres (the thermal equilibrium is impossible due to the dissipation). The density however is inhomogeneous: the inelastic cooling opens the possibility of having stable inhomogeneous spatial profiles of density, which is absent in elastic systems. 

This possibility of having states of granular systems with inhomogeneous stationary profile of density (called below the "IS" for "inhomogeneous states") was first discovered in \cite{MFV,IF}. In the IS 
the heat flux caused by the inhomogeneity of temperature is balanced by the inelastic cooling which is also inhomogeneous so that the spatial profile of the temperature is 
preserved up to the depletion of overall amplitude. This spatial structure stably exists when the entropy monotonously decreases to $-\infty$ (the decrease stops eventually when the inner degrees of freedom start to return energy to the translational ones at the collisions), cf. \cite{Brey0,Baldassarri,Puri}. However, the consideration of the thermodynamic limit of the IS, presenting the highest interest, is not possible within the dilute limit: it was found that the maximal density of the IS grows exponentially with the system size, so the density becomes comparable with the density of dense packing for large size of the system (thermodynamic limit). Thus the study of the thermodynamic limit necessitates considering dense regions. There, usually, the study can be performed only phenomenologically due to the strong coupling between the particles (for example, even the precise equation of state is unknown for liquids). 

We succeed in deriving the dense IS that hold in the thermodynamic limit without approximations. That is, the IS are derived for granular liquids, taking into account the finite size of the particles and the corresponding excluded volume effects. The derivation uses description that works uniformly both for gas and liquid phases including the possibility of coexistence of the phases. 
This is despite that the constitutive relations that appear in the fluid mechanical equations (such as the equation of state) are not known explicitly in the liquid phase. We note that the particular scaling form of those functions, that holds due to the absence of the intrinsic energy scale for hard spheres, admits the IS solutions. Due to the robustness of this observation, one can expect that the existence of the IS is a fundamental property of granular materials.

The IS have unusual properties. These are states of coexistence of liquid and gas phases. The phases are described by the time-independent density profile that varies continuously in space from large values in the liquid phase to low values in the gaseous phase.

Probably the most striking property of the IS is its particular organization of particles in space due to which the system does not dissipate energy in the thermodynamic limit. This is not because the particles freeze: the temperature is finite. However the flow of energy in space due to the inhomogeneity of temperature compensates the local energy losses due to inelastic collisions so that the energy obeys $[1+t/t_c]^{-2}$ where $t_c$ diverges in the thermodynamic limit. In particular, since the local energy dissipation rate is determined completely by the local density and temperature, this signifies that these fields "know" of the system size in the IS. 

This is in sharp contrast to the well-known homogeneous cooling states, HS, \cite{Haff}, where the decay obeys the same law $[1+t/t_c]^{-2}$, but $t_c$ is a local property independent of the size of the system. 

Furthermore, the total entropy decreases due to the dissipation logarithmically in time. This decrease is proportional to the number of collisions that occurred in the system. Thus, roughly, the entropy decreases in each collision by the same, time-independent amount.

Our derivation of the IS holds for systems where on average the system is dilute, so the fraction of space occupied by the gas phase is close to unity. (This is, in particular, the case of the clustering instability of the HS of granular gases, see below). However, the small fraction of space occupied by the liquid contains the fraction of the total mass that is close to unity. In contrast, the energy of the system is predominantly contained in the gaseous state, where the particles' speed is higher. Thus in the IS both phases are present significantly, one carrying the mass, the other the energy of the total system.

Further insight into the physics of the IS is reached by noting that the IS becomes a "time"-independent solution, when considered in the natural time variable, the number of collisions, using time-rescaled fields.  To demonstrate this we provide a time-dependent transformation that transforms the original equations, that do not have explicit time-dependence, into the equations that do not have explicit time-dependence too. The IS is the stationary solution in rescaled variables. We use this transformation to demonstrate that the evolution of small perturbations of the IS obeys the power-law behaviour in time.

Once the IS solutions are derived, the question is if they are stable. The study of this question in the dilute case has a long history. It started from the study of the stability of the HS \cite{Haff}. This state plays the role of the equilibrium state for systems with size smaller than an intrinsic instability length $l_{cr}$ times $\pi$, which relax to the HS universally at large times. The scale $l_{cr}$ is of order of the mean free path over $\sqrt{1-r}$. However the HS is unstable for larger systems due to the famous clustering instability \cite{Haff,Goldhirsch1,Goldhirsch3,McNamara1,McNamara2}. That instability, obtained by linearizing the equations around the HS, demonstrated the formation of clusters in the system that spontaneously break the translational symmetry. It was shown in \cite{MFV,IF} that when the system size passes the instability length (supercritical systems), simultaneously with uniform solutions getting unstable, the IS appear (that do not exist at smaller length). Based on the analogy with instability in the Newtonian fluids, one can expect that the IS plays the role of the HS for supercritical systems and constitutes the result of evolution at large times. This was shown to be the case for not too large systems for the channel geometry of the particles' container \cite{MFV}. 

In the latter case the fluid mechanical fields depend only on the coordinate along the channel \cite{ELM,MP,Fouxon1,Fouxon2,Puglisi}.
The limit of fast sound was considered in \cite{MFV}, where the sound propagation time $t_s$ across the system is much smaller than the characteristic time of the decay of energy due to inelasticity. Then, since $t_s$ is the characteristic time of relaxation of the pressure to the equilibrium uniform value, the inelastic evolution happens on the background of the pressure that is constant in space (but not in time). 
As a result, one can  reduce the compete system of fluid mechanical equations to one integro-differential scalar equation. The numerical study of that equation demonstrated that the IS are stable. Furthermore, the IS provide universal result of the long-time evolution of the granular gas. In particular, they determine the ultimate result of the clustering instability of the HS \cite{Haff,Goldhirsch1,Goldhirsch3,McNamara1,McNamara2}. 

Thus the IS provided the first consistent prediction on the long-time result of the non-linear development of the instability in a certain limit. This limit, though, demands that the length of the channel is not too large, so it cannot be used to study the thermodynamic limit. The consideration of the latter limit demands considering the full system of the fluid mechanical equations, and it was not performed previously. 

In this work we perform the numerical simulation of the complete system of the fluid mechanical equations of the dilute granular gas in the channel (so that the integral equation simulated in \cite{MFV} is a reduction of this system in the limit of moderate system size). We demonstrate that the IS is the result of the long-time evolution of the system for arbitrary length of the channel. Since for arbitrarily large but fixed size of the system, the dilute gas holds in the limit of small size of the particles, our result signifies that the IS is the result of the long-time evolution of the system in the thermodynamic limit taken {\it after} the limit of zero particle size. 

To complete the proof that the IS is the universal result of the long time evolution of the collection of inelastic hard spheres in the channel, one has to deal with the finite particles' size. The maximal density on the IS of the dilute gas grows exponentially with the size of the system, so the consideration of the thermodynamic limit of the collection of (real) finite-size particles necessitates the study of the fluid-mechanical equations not constrained by the diluteness assumption. We perform this study using the following consideration.

Our derivation of the IS does not assume that the fluid is dilute. Though the form of the coefficients of the fluid mechanical equations is not known in the dense case, we demonstrate that the structure of the solution can be determined independently of that form if the system is dilute on average (that is if the spheres were distributed uniformly they would form the dilute gas). 
The IS in this case consists of the gas phase that occupies the volume's fraction close to unity and the liquid phase that occupies the remaining volume. The liquid phase produces effectively a wall boundary condition for the gas where the wall is located at the boundary between the phases. 

This structure of the IS in the dense case is indicating strongly that it is stable. Indeed, the gas phase is locally stable in view of the stability of the dilute IS that was proved numerically. The liquid phase is locally stable too because the excluded volume effects do not allow significant growth of the density in the liquid state. Thus the IS is stable locally. If to discard the unlikely possibility of non-local mechanism of instability (note that the mechanism of instability of the HS is local \cite{Haff,Goldhirsch1,Goldhirsch3,McNamara1,McNamara2}), then this completes the demonstration that the IS is the universal result of the long-time evolution in the channel.

The IS solutions considered in this work depend on one spatial coordinate only (we stress that microscopically the system is fully three or two-dimensional), describing the physics of granular materials in channels. The IS  solutions that depend on two or three variables exist too, so it is natural to pose the question of their role in situations where the fluid mechanical flows depend on two or three coordinates, which is the situation of box geometry. Since the physics of the IS solutions consists of the balance of the Laplacian term describing heat conduction and local non-linearity corresponding to the inelastic cooling, then the role of dimension should be important because it is important for the Laplacian. Recent numerical results indicate that in higher dimensions the IS are unstable. The work of \cite{Kolvin} reports that two-dimensional fluid-mechanics of granular gases produces finite-time singularities of density. This increases the importance of considering the dense IS introduced in this work. Indeed, finite particles' size effects will necessarily regularize the singularity of the density, cf. \cite{Puglisi}. The corresponding study if the dense IS is the result of the long-time evolution of supercritical systems in the higher-dimensional case is the topic for future work. 

It is assumed in our study that the initial conditions do not contain the solid state phase and that the solid phase does not form in the system as a result of the evolution. This is reasonable due to the growth of the pressure in the liquid state. This guarantees the applicability of the fluid mechanical description. It is to be stressed that the fluid mechanical equations that we use are valid both in the dilute (gaseous) and in the dense (liquid) phases of the matter. Though the form of the coefficients (viscosity, heat conductivity, equation of state...) is not known precisely in the dense region, we succeed in dealing with the equations based on the separability of the dependence of the coefficients on the temperature and density that holds for hard-core particles. The result applies to the two-dimensional fluid of hard disks too. 

The following text is structured as follows. In the next Section we derive the general form of the fluid mechanical equations of the hard spheres that collide inelastically. We show that these equations have a particular scaling form that holds due to the absence of energy scale in the the hard-spheres' interaction. In Section \ref{in} we derive the IS solutions of the fluid mechanical equations and demonstrate their basic properties (that do not depend on the dimension). The following Section introduces the time-dependent transformation of the fluid mechanical equations that transforms the IS to the time-independent solution of a system with no explicit time-dependence, implying power-law behavior of the perturbations of the IS. In Section \ref{st} we provide the implicit form of the IS that depends on one spatial coordinate in terms of coefficients of the dense fluid. The next Section provides the qualitative structure of the solution, that is relevant for the later consideration of the dense case. The following Section describes the low-density limit of the IS not confined to the case of the dependence on one coordinate. The study of the IS of the dilute gas that depend on one coordinate only is performed in Section \ref{s7}. Though this case was solved previously \cite{MFV,IF}, we provide the solution to stress its thermodynamic limit and to make the paper self-contained. The IS in the dilute case serve the basis for the study of the IS in the dense case in Section \ref{dense}. Section \ref{s8} deals with introducing the Lagrangian form of the fluid mechanical equations that is considerably more suitable for the numerical simulation. 
The latter is reported in Section \ref{numerics}. It is shown that the IS provide the result of the long-time evolution of initial conditions that are arbitrary. Section \ref{finite} discusses the recent conjecture on the finite-time singularity of the density of the dilute gas in the light of the dense IS obtained here. 
Finally, in the Conclusion we discuss the implications of our results on the general study of the granular materials. Further questions risen by our results are considered.

\section{Fluid mechanics of dense collections of inelastically colliding hard spheres}

The fluid of elastic hard spheres has an exceptional feature that in the equilibrium the only energy scale is the temperature $T$ (in usual fluids there is an energy scale that characterizes the interactions). Thus, the intensive quantities, like the pressure $p$ or the thermal conductivity $\kappa$ (determined by the equilibrium properties via the Kubo formula) are determined completely by the density $\rho$, the particles' diameter $\sigma$ and $T$ (we set the particles' mass equal to one with no loss). The only one of those quantities that contains the units of time is $T$. This allows to use the dimensional analysis to determine uniquely the dependence of the intensive quantities on $T$. In this way one fixes the temperature dependence in the constitutive equations for the functions that appear in the fluid mechanical equations.

The inelasticity described by the dimensionless constant coefficient of the normal restitution $r$ does not introduce a new energy scale. This is a unique property of the considered model of the inelasticity, while other models typically contain an energy scale characterizing the inelastic part of the interactions. Thus, the equations of hydrodynamics of the granular fluid of hard spheres also have coefficients which dependence on $T$ is fixed for any density.

We pass to the description of the fluid mechanics of inelastic hard-sphere fluids. This form holds both for the dilute gases and for the dense fluids. We assume that inelasticity is small, i. e. $1-r\ll 1$. Only under this assumption one can use the fluid mechanics, at least in its traditional form, that assumes the local thermal equilibrium. When $1-r\ll 1$
the local relaxation to equilibrium that happens within a local relaxation time $\tau_{rel}$ is weakly influenced by the inelasticity. In particular, the inelastic energy depletion that happens within $\tau_{rel}$ is small.  Further evolution of the system consists of the evolution of the fields that describe the variation of the parameters of the local thermal equilibrium in space, i. e. the density $\rho(\bm x, t)$, the velocity $\bm v(\bm x, t)$ and the temperature $T(\bm x, t)$. The evolution of these fields is derived from the local balance of mass, momentum and energy. The equations are obtained by expressing the latter and their currents via $\rho(\bm x, t)$, $\bm v(\bm x, t)$ and $T(\bm x, t)$. The inelasticity produces a local term in the energy equation that describes inelastic energy losses.

We pass to the description of the equations. The inelasticity does not change the local laws of conservation of mass and momentum. Using isotropy and Galilean invariance one finds the following general form of the fluid mechanical equations \cite{Landau,Landau10}
\begin{eqnarray}&&\!\!\!\!\!\!
\frac{\partial\rho}{\partial t}+\nabla\cdot[\rho \bm v]=0,\ \ \rho\left[\frac{\partial v_i}{\partial t}+(\bm v\cdot\nabla) v_i
\right]=-\nabla_i p+\frac{\partial \sigma_{ik}}{\partial x_k},\nonumber \\&&
\sigma_{ik}=\eta\left(\frac{\partial v_i}{\partial x_k}+\frac{\partial v_k}{\partial x_i}-\frac{2}{d}\delta_{ik}\nabla\cdot \bm v\right)+\zeta\delta_{ik}\nabla\cdot \bm v, \label{constf}
\end{eqnarray}
where $\sigma_{ik}$ is the viscous momentum flux (cf. \cite{Landau}), $d=2$ corresponds to the case of hard discs and $d=3$ to the case of hard balls.
Using the dimensional analysis we have the following general form of the intensive quantities $p$, $\eta$ and $\zeta$,
\begin{eqnarray}&&\!\!\!\!\!\!\!\!\!\!\!\!
p=T F_1(\rho , r),\ \ \eta=\sqrt{T}F_2(\rho , r),\ \ \zeta=\sqrt{T}F_3(\rho , r), \label{const}
\end{eqnarray}
where $F_i(\rho, r)$ are certain functions. Though this is not necessary to demonstrate the IS, we use $1-r\ll 1$ to set $r=1$ in the
equations above. It will be clear below that to the leading order in $1-r$ the difference of $r$ from $1$ needs to be accounted only in the energy loss term. The reason is that this is the only term for which there is a degeneracy at $r=1$, where it vanishes identically. Its smallness is compensated by the large time of evolution (however small, the inelasticity will produce the final state which is completely different from the $r=1$ case). The rest of the terms in the equations have a finite limit at $r\to 1$, so their effect is perturbative. Below we define $F_i(\rho)\equiv F_i(\rho, r=1)$. 

The function $F_1(\rho)$ gives the pressure of the fluid of elastic hard-spheres and can be expressed with the help of the free energy $F_{free}$ of
that fluid. For $N$ particles,
\begin{eqnarray}&&
F_{free}=F_{id}+N T F(\rho),
\end{eqnarray}
where $F_{id}$ is the free energy of the ideal gas. The function $F(\rho)$ is determined by the configuration integral and its meaning is that it gives the deviation of the entropy per particle $s$ from the one of the ideal gas,
\begin{eqnarray}&&
s=-\frac{1}{N}\left(\frac{\partial F_{free}}{\partial T}\right)=s_{id}-F(\rho),
\end{eqnarray}
where $N$ is the number of spheres and $s_{id}=\ln[T^{1/(\gamma-1)}/\rho]+const$ is the entropy per particle for the ideal gas.
The function $F(\rho)$ vanishes in the limit $\rho\to 0$ and its Taylor expansion starts from the linear term. For the equation
of state one obtains
\begin{eqnarray}&&
p=\frac{\rho^2}{N}\left(\frac{\partial F_{free}}{\partial \rho}\right)=\rho T+\rho ^2F'(\rho)T,
\end{eqnarray}
Thus we have
\begin{eqnarray}&&
F_1(\rho)=\rho+\rho ^2F'(\rho).
\end{eqnarray}
To write down the remaining fifth equation of the fluid mechanics, we first write the equation for the usual fluid of elastic hard spheres and then introduce the energy loss term due to inelasticity. The energy balance equation can be written in the form \cite{Landau}
\begin{eqnarray}&&
\rho T\left(\frac{\partial}{\partial t}+\bm v\cdot \nabla \right) s
=\sigma_{ik}\frac{\partial v_i}{\partial x_k}+\nabla\cdot[\kappa\nabla T],\label{entropy}
\end{eqnarray}
where $\kappa$ is the thermal conductivity. One can write $\kappa=\sqrt{T}F_4(\rho)$ similarly to Eqs.~(\ref{const}). 

For the fluid of inelastic hard spheres the pressure has a particular role. The fluid tends to make the pressure uniform which makes it a
convenient variable to work with. To pass from $s$ to $p$ we insert $\rho T=p/\left[1+\rho F'(\rho)\right]$ into the expression for the entropy of the ideal gas $s_{id}$ which gives
\begin{eqnarray}&&
s=\frac{\ln p}{\gamma-1}-\frac{\ln \left[1+\rho F'(\rho)\right]}{\gamma-1}-\frac{\gamma\ln \rho}{\gamma-1}
-F(\rho).
\end{eqnarray}
Putting this 
into Eq.~(\ref{entropy}), with $\gamma=5/3$ for three-dimensional fluid of hard balls and $\gamma=2$ for the two-dimensional fluid of hard discs and using the continuity equation, one finds
\begin{eqnarray}&&
\left(\frac{\partial}{\partial t}+\bm v\cdot \nabla \right)p=
-F_5(\rho) p\nabla\cdot \bm v+\frac{(\gamma-1)F_1(\rho)}{\rho}
\nonumber\\&&\times
\Biggl(\sigma_{ik}\frac{\partial v_i}{\partial x_k}+\nabla\cdot[\sqrt{T}F_4(\rho)\nabla T]\Biggr), \label{pres}
\end{eqnarray}
where $F_5(\rho)$ is defined by
\begin{eqnarray}&&\!\!\!\!\!\!\!\!\!\!\!\!\!\!\!\!\!\!
F_5(\rho)=\frac{\gamma+2\gamma\rho F'(\rho)+\rho^2 F''(\rho)+(\gamma-1)\rho^2 F'^2(\rho)}{\left[1+\rho F'(\rho)\right]}.\label{const}
\end{eqnarray}
To include inelasticity one should introduce into the equation the energy loss term which 
form can be fixed by dimensional analysis. We obtain 
\begin{eqnarray}&&
\left(\frac{\partial}{\partial t}+\bm v\cdot \nabla \right)p=
-F_5(\rho) p\nabla\cdot \bm v-\Lambda(\rho)\rho^{1/2}p^{3/2}
\nonumber\\&&
+\frac{(\gamma-1)F_1(\rho)}{\rho}\nabla\cdot\left[\sqrt{\frac{p}{F_1(\rho)}}F_4(\rho)\nabla \left(\frac{p}{F_1(\rho)}\right)\right]
\nonumber\\&&
+\frac{(\gamma-1)F_1(\rho)}{\rho}\sigma_{ik}\frac{\partial v_i}{\partial x_k},
\label{fluideq}
\end{eqnarray}
where $\Lambda(\rho)$ tends to a positive constant $\Lambda$ in the dilute limit \cite{BP}. The system of Eqs.~(\ref{constf}),(\ref{fluideq}) is the complete system of equations of the granular fluid of hard balls. 
This has a special dependence on the fields of $p$ and $\rho$, where both the cooling and the thermal conductivity terms depend
on $p$ as $p^{3/2}$. It turns out that based on this form only one can find a new type of solutions, compared to the usual fluid of elastic hard spheres, that hold due to the inelasticity.

\section{Inhomogeneous Solutions}

We look for the solution to Eqs.~(\ref{constf}),(\ref{fluideq}) that obeys $\bm v\equiv 0$. The continuity equation implies that these solutions
describe a stationary distribution of particles in space, $\rho=\rho(\bm x)$. The momentum equation gives that the pressure must be spatially uniform and it can depend only on time, $p=p(t)$. Finally, equation (\ref{fluideq}) under the assumption $\rho=\rho(\bm x)$ and $p=p(t)$ can be written as follows
\begin{eqnarray}&&\!\!\!\!\!\!\!\!\!\!\!\!\!\!
\frac{1}{p^{3/2}}\frac{dp}{dt}\!=\!-\!\Lambda(\rho)\rho^{1/2}
\!+\!\frac{(\gamma\!-\!1)F_1(\rho)}{\rho}\nabla\!\cdot\!\left[\frac{{\tilde F}_4(\rho)}{\sqrt{\rho}}\nabla \frac{1}{\rho}\right],\label{red}
\end{eqnarray}
where we defined ${\tilde F}_4(\rho)\equiv F_4(\rho)F_1'(\rho)\rho^{5/2}/F_1^{5/2}(\rho)$.
Since the LHS of Eq. (\ref{red}) by assumption is a function of time, while the RHS is the function of coordinate, then the solutions exist if both sides are equal to
a constant $-c$, where minus is written for further convenience,
\begin{eqnarray}&&\!\!\!\!\!\!\!\!\!\!\!\!\!\!\!
\frac{1}{p^{3/2}}\frac{dp}{dt}=-c, \nonumber \\&&\!\!\!\!\!\!\!\!\!\!\!\!\!\!\!
\frac{(\gamma-1)F_1(\rho)}{\rho}\nabla\cdot\left[\frac{{\tilde F}_4(\rho)}{\sqrt{\rho}}\nabla \left(\frac{1}{\rho}\right)\right]-\Lambda(\rho)\rho^{1/2}=-c.\label{eq00}
\end{eqnarray}
Dividing the above equation by $F_1(\rho)/\rho$ and integrating over space we find
\begin{eqnarray}&&
c=\frac{\langle \Lambda(\rho)\rho^{3/2}F_1^{-1}(\rho)\rangle}{\langle \rho F_1^{-1}(\rho)\rangle},\ \
\langle f\rangle \equiv \frac{1}{\Omega}\int_{\Omega} f(\bm x)d\bm x,\nonumber
\end{eqnarray}
where $\Omega$ is the system volume, the angular brackets stand for spatial averages and 
we assumed that either the heat flux vanishes at the boundary or that the periodic boundary conditions hold, so the boundary terms vanish. 
We obtain that the pressure obeys the power-law 
\begin{eqnarray}&&\!\!\!\!\!\!\!\!\!\!\!\!\!\!\!
p(t)\!=\!\frac{p(0)}{\left[1+t/t_c\right]^2},\ \ \! t_c\!\equiv\! 
\frac{2\langle \rho F_1^{-1}(\rho)\rangle}{\langle \Lambda(\rho)\rho^{3/2}F_1^{-1}(\rho)\rangle p^{1/2}(0)}, \label{pres}
\end{eqnarray}
where the density field obeys the non-linear PDE
\begin{eqnarray}&&\!\!\!\!\!\!
\frac{(\gamma-1)F_1(\rho)}{\rho}\nabla\cdot\left[\frac{{\tilde F}_4(\rho)}{\sqrt{\rho}}\nabla \left(\frac{1}{\rho}\right)\right]
-\Lambda(\rho)\rho^{1/2}
\nonumber\\&&
=-\frac{\langle \Lambda(\rho)\rho^{3/2}F_1^{-1}(\rho)\rangle}{\langle \rho F_1^{-1}(\rho)\rangle}.\label{eq}
\end{eqnarray}
Introducing 
\begin{eqnarray}&&\!\!\!\!\!\!
K(\rho)\equiv -\int \frac{{\tilde F}_4(\rho) d\rho}{\rho^{5/2}},\ \ \nabla K=\frac{{\tilde F}_4(\rho)}{\sqrt{\rho}}\nabla \left(\frac{1}{\rho}\right),\label{coor}
\end{eqnarray}
one can rewrite Eq.~(\ref{eq}) in the form 
\begin{eqnarray}&&
\nabla^2 K=f\left[\rho(K)\right],
\label{lap}
\end{eqnarray}
where $\rho(K)$ is the inverse function of $K(\rho)$ and
\begin{eqnarray}&&
f(\rho)\equiv {\tilde F}(\rho)-\frac{\langle  {\tilde F}(\rho)\rangle}{\left[1+\rho F'(\rho)\right] \langle \left[1+\rho F'(\rho)\right]^{-1}\rangle}
\nonumber\\&&
{\tilde F}(\rho)\equiv \frac{\Lambda(\rho)\rho^{3/2}}{F_1(\rho)(\gamma-1)}. 
\end{eqnarray}
The previous work \cite{MFV,IF} characterized the solutions in the limit of the dilute gas where one can neglect the term with $F'$ and ${\tilde F}(\rho)\approx \Lambda \rho^{1/2}/(\gamma-1)$ [where $\Lambda=\Lambda(\rho=0)$]. One finds $f(\rho)\approx  \Lambda \left[\langle \rho^{1/2} \rangle-\rho^{1/2}\right]/(\gamma-1)$. Solutions in the dilute case that depend on one coordinate only were worked out in detail in  \cite{MFV}, while the higher dimensional case for spherically symmetric situation in was considered in \cite{IF}. It is clear that there are non-spherically symmetric solutions that depend on two or three coordinates, however the discussion of these solutions is beyond the scope of this work. 

Studying the solutions in the dense case, one can consider the model equation of state $F'(\rho)=\pi g(\rho)/\sqrt{3}$ of Carnahan and Starling \cite{CS} where
\begin{eqnarray}
g(\rho)=\frac{1-7\pi \rho/[32\sqrt{3}]}{(1-\pi \rho/[2\sqrt{3}])^2}. \label{CS}
\end{eqnarray}
is the equilibrium pair correlation function of hard disks
at contact (we use here units with $\sigma=1$ and consider $d=2$, the study of $d=3$ is similar). The corresponding expression for the cooling coefficient $\Lambda(\rho)$ derived in \cite{jenkins} in the spirit of Enskog theory is $\Lambda(\rho)=\Lambda g(\rho)$. Thus 
\begin{eqnarray}
{\tilde F}(\rho)=\frac{\Lambda g(\rho)\rho^{1/2}}{(\gamma-1) \left[1+\pi\rho g(\rho)/\sqrt{3}\right]}. \label{tildeF}
\end{eqnarray}

The simplest solution is the uniform solution $\rho=\rho_0$ where $\rho_0$ is a constant,
\begin{eqnarray}&&\!\!\!\!\!\!\!\!\!\!\!\!\!\!\!\!\!\!\!
p(t)=\frac{p(0)}{\left[1+t/t_c\right]^2},\ \ t_c^{unif}= 
\frac{2}{\Lambda(\rho_0)\rho_0^{1/2} p^{1/2}(0)}. \label{uniformly cooling state}
\end{eqnarray}
This solution is the well-known HS in the dilute gas limit $\rho_0 \sigma^3\ll 1$, where $\Lambda(\rho_0)$ tends to a constant $\Lambda\neq 0$, see \cite{Haff}.
However, we could not find in the literature these solutions in the dense case, where $\rho_0 \sigma^3\sim 1$ and 
$\Lambda(\rho_0)$ differs from $\Lambda$ appreciably. This solution is quite notable because $\Lambda(\rho)=\Lambda g(\rho)$ diverges when the density approaches the density of the dense packing, see Eq. (\ref{CS}), so that the cooling becomes infinitely fast.  
Further, while the
instability of the uniform solutions for the gases is well-known, see \cite{Goldhirsch1,Goldhirsch3,McNamara1,McNamara2}, there is no corresponding study for the dense fluid. It is clear though that the instability of perturbations with a large enough wave-length (which existence demands the system size to be sufficiently large), that was proved for the gases, holds for dense fluids too, because the instability mechanism does not depend on the diluteness \cite{Goldhirsch1,Goldhirsch3,McNamara1,McNamara2,Brey}. The study of the dependence of the critical length on $\rho_0$ in the dense case is left for future work. Here we confine the consideration to the dilute case where the total number of particles is such that $N\sigma^3/\Omega\ll 1$, which guarantees that the uniform state is unstable. 

Thus for large system size the unstable uniform solutions have little practical value. However, 
there are also inhomogeneous solutions to Eq.~(\ref{lap}), see \cite{IF}. These can provide the final state of the fluid at large system size, see below and \cite{MFV}, giving importance to their consideration. 
Before we discuss those solutions we consider the behavior of the energy and the entropy on the IS, that can be found independently of the form of the density.

The thermal energy density is equal $\rho T$ (the one of ideal gas), so that the total system's
energy $E(t)$ obeys 
\begin{eqnarray}&&
E(t)=p(t)\int \frac{d \bm x}{1+\rho F'(\rho)}=\frac{E(0)}{\left[1+t/t_c\right]^2},
\end{eqnarray}
where we used that the integral is a constant that can be fixed using the initial value of $E(t)$. 
We used that the system has no macroscopic kinetic energy. While the energy decay is necessary, whether the entropy decays or grows 
is less obvious. The energy losses cause the entropy to decrease, but the spatial inhomogeneity leads to the increase of entropy. We use that
$s=\ln[p^{1/(\gamma-1)}]$ plus a function of the density. Since the density in the IS is time-independent, then the entropy is
\begin{eqnarray}&&\!\!\!\!\!\!
S(t)=\int \frac{\rho \ln p}{\gamma-1}d\bm x+B=\frac{N \ln p(t)}{\gamma-1}+B
\nonumber\\&&
=S(0)-\frac{2N \ln \left[1+t/t_c\right]}{\gamma-1},
\end{eqnarray}
where $B$ is a constant determined by $S(0)$. Thus the IS are the states for which the entropy decreases
logarithmically to minus infinity as $t$ grows. The system continuously organizes with chaotic disorder decreasing due to the energy decrease.
Of course, as the fluid cools down, eventually the effective description breaks down as a physically realistic description so the physical entropy
stays finite (when the inner degrees of freedom have energy comparable with the one of the translational degrees of freedom they will stop to be the energy sink described by inelasticity, but rather will exchange energy with the translational degrees of freedom).

Finally we consider the total number of particles' collisions $N_c(t)$ that occurred by the time $t$. By dimensional analysis, the local rate of collisions $\Gamma(t)$ obeys $\Gamma(t)=L(\rho)p^{1/2}$, where $L(\rho)$ is a function of $\rho$. We find
\begin{eqnarray}&&\!\!\!\!
\frac{dN_c}{dt}=\int L(\rho)p^{1/2} d\bm x=\frac{const}{1+t/t_c},
\end{eqnarray} 
which simply says that the local rate of collisions is proportional to the typical relative velocity of the particle $T^{1/2}$. Integrating, 
\begin{eqnarray}&&
N_c(t)=const \times t_c\ln[1+t/t_c].
\end{eqnarray} 
The number of collisions grows only logarithmically in time. The particles collide more and more rarely with time due to the decrease of their velocity. 

It is observed that the decrease of the entropy is proportional to the number of collisions that occurred in the system,
\begin{eqnarray}&&\!\!\!\!\!\!
S(t)-S(0)=-\frac{2N N_c(t)}{const \times t_c(\gamma-1)},
\end{eqnarray}
 Thus on the IS the entropy decreases by roughly the same quantity in each collision.

We showed in this section that the fluid mechanical equations of dense granular fluids of hard spheres have exact solutions for which the pressure is spatially uniform and the generally inhomogeneous profile of the density exists in space. The solutions hold due to the special combination of scalings where the pressure is proportional to $T$, while both the cooling and the heat conduction scale as $T^{3/2}$. The density profile is preserved via 
the balance of inhomogeneous cooling and heat conduction which sum to a spatially-independent value.
Due to the robustness of this balance we expect that the IS are general and present also when $1-r$ is not small and the fluid mechanics does not hold.

The decay of the pressure and of the energy in the IS obeys a universal power-law with exponent two independently of the details of the density profile. The entropy decreases logarithmically in time and proportionally to the number of particles' collisions that occurred in the system. The number of collisions $N_c(t)$ turns out to be the natural time variable in which to consider the evolution, as we pass to show.  
Below we use the arbitrariness in the choice of the initial moment of time to set $p(0)=1$ in the IS.

\label{in}
\section{Behavior of small perturbations of the IS}

The system of equations (\ref{constf}),(\ref{fluideq}) admits a time-dependent transformation of variables such that in the new variables
the system is still time-independent.  The unique property of this transformation is that the IS become time-independent in the new variables. In particular, the transformation allows us to conclude that the linearized
perturbations around the IS have a power-law behavior in time.

For an arbitrary constant $C$, we pass to the  new functions $\rho'$, $p'$ and $\bm v'$ defined by
\begin{eqnarray}&&
p=\frac{p'}{\left[1+C t\right]^2},\ \ \rho=C^2 \rho',\\&& \bm v=\frac{\bm v'}{C\left[1+Ct\right]},\nonumber
\end{eqnarray}
and the new space and time variables
\begin{eqnarray}&&
\tau=\ln\left[1+C t\right],\ \ \frac{\partial}{\partial t}=\frac{C}{1+Ct}
\frac{\partial}{\partial \tau},\nonumber\\&&
\bm x'=C^2 \bm x.
\end{eqnarray}
We have
\begin{eqnarray}&&
\frac{\partial \bm v}{\partial t}=\frac{1}{\left[1+C t\right]^2}
\frac{\partial \bm v'}{\partial \tau}-\frac{\bm v'}{\left[1+C t\right]^2},\\&&
\frac{\partial p}{\partial t}=\frac{C}{\left[1+C t\right]^3}\frac{\partial p'}{\partial \tau}-
\frac{2Cp'}{\left[1+C t\right]^3}.
\end{eqnarray}
In the new variables the system (\ref{constf}),(\ref{fluideq}) becomes
\begin{eqnarray}&&
\frac{\partial\rho'}{\partial \tau}+\nabla'\cdot[\rho' \bm v']=0,\ \ \rho'\left[\frac{\partial v'_i}{\partial \tau}-\bm v'+(\bm v'\cdot\nabla') v'_i
\right]\nonumber\\&&=-\nabla'_i p'
+\frac{\partial \sigma'_{ik}}{\partial x'_k}
,\ \ \ \ 
\frac{\partial p'}{\partial \tau}-2p'+\bm v'\cdot \nabla'p'\nonumber\\&&
\!\!\!\!\!\!=
-F_5(C^2\rho')
p'\nabla'\cdot \bm v'-\Lambda(C^2\rho')\rho'^{1/2}p'^{3/2}
\nonumber\\&&\!\!\!\!\!\!
+F_6(C^2\rho')\Biggl(\sigma'_{ik}\frac{\partial v'_i}{\partial x'_k}
+\nabla'\cdot\biggl[\sqrt{\frac{p'}{\rho'\left[1+\rho'F'(\rho')\right]}}
\nonumber\\&&
F_4(C^2\rho')\nabla' \left(\frac{p'}{\rho'\left[1+\rho'F'(\rho')\right]}\right)\biggr]\Biggr).\nonumber
\end{eqnarray}
where
\begin{eqnarray}&&\!\!\!\!\!\!
\sigma'_{ik}=\sqrt{\frac{p'}{\rho'\left[1+\rho'F'(\rho')\right]}}F_2(C^2\rho')\biggl(\frac{\partial v'_i}{\partial x'_k}+\frac{\partial v'_i}{\partial x'_k}
\nonumber\\&&\!\!\!\!\!\!
-\frac{2}{3}\delta_{ik}\nabla'\cdot \bm v'\biggr)
+\sqrt{\frac{p'}{\rho'\left[1+\rho'F'(\rho')\right]}}F_3(C^2\rho')\delta_{ik}\nabla'\cdot \bm v'.\nonumber
\end{eqnarray}
Importantly, the IS solution is time-independent in the new variables and it reads
\begin{eqnarray}&&
\rho'(\bm x')=\frac{1}{C^2}\rho_0\left(\frac{\bm x'}{C^2}\right),\ \ C=\frac{c}{2},\ \ 
\bm v'=0,\nonumber\\&&
p'^{-1/2}=\frac{\langle F_6^{-1}(C^2\rho')\Lambda(C^2\rho')\rho'^{1/2}\rangle}{2\langle F_6^{-1}(C^2\rho')\rangle}=const,
\end{eqnarray}
as can be verified from Eq.~(\ref{eq}) on $\rho_0$. It should be noticed that though the transformation allows for any value of $C$, different values of $C$ do not generate new solutions. Rather they describe the generation of the new solutions using the translational invariance in time: the IS remain the solutions if $t$ is changed to $t$ plus a constant. 
For example, for $C=1$, one recovers the previous solution with $p^{1/2}(0)$ such that $t_c=1$.

The described transformation is useful for studying the behavior of small perturbations of the IS. There one can choose $C=1/t_c$, so that $\tau(t)=(\gamma-1)\left[S(0)-S(t)\right]/(2N)$ is proportional to the change of the entropy.  In the new variables the eigenmodes of the linearized operator that describes the behavior of small pertbatioins near the IS behave exponentially in time. Thus the behavior of the modes is a power-law in the physical time. 

Clearly,  $\tau(t)$ can also be taken as the number of collisions $N_c(t)$. 
Thus the IS are time-independent in the rescaled fields when considered as a function of the number of collisions that occurred since the initial moment of time. 

\label{transformation}

\section{IS depending on one coordinate}
\label{st}

The density field in the IS obeys the non-linear PDE (\ref{lap}) [or (\ref{eq})] that must be solved either with von Neumann boundary conditions (that describe the demand that the heat does not flow through the boundary) or with periodic boundary conditions. It is not possible to find the solution without specifying the form of $f(K)$ except for the case where the density depends on one spatial coordinate $x$ only. In this case  Eq.~(\ref{lap}) describes one-dimensional mechanical motion where $f(K)$ is the force (cf. the radially symmetric solution provided in \cite{IF}), 
\begin{eqnarray}&& \!\!\!\!\!\!
\frac{d^2 K}{dx^2}=-\frac{\partial U\left[\rho(K)\right]}{\partial K} =-\frac{\rho^{5/2}\partial_{\rho} U}{{\tilde F}_4(\rho)}
\left[\rho=\rho(K)\right]
;\label{isaac}\\&& \!\!\!\!\!\!
\partial_{\rho} U =-\frac{{\tilde F}_4(\rho)f(\rho)}{\rho^{5/2}}
 =\frac{{\tilde F}_4(\rho){\tilde F}(\rho)}{\rho^{5/2}}
-\frac{{\tilde F}_4(\rho)\langle  {\tilde F}(\rho)\rangle}{\rho^{3/2} F_1(\rho) \langle \rho/F_1(\rho)\rangle}.
\nonumber\end{eqnarray}
One has 
\begin{eqnarray}&& \!\!\!\!\!\!\!\!\!\!\!\!
U=\int \frac{{\tilde F}_4(\rho){\tilde F}(\rho)}{\rho^{5/2}}d\rho-
\frac{\langle  {\tilde F}(\rho)\rangle}{\langle \rho/F_1(\rho)\rangle} \int 
\frac{{\tilde F}_4(\rho) d\rho}{\rho^{3/2} F_1(\rho)}. \label{pot}
\end{eqnarray}
It follows that the "energy" 
\begin{eqnarray}&& \!\!\!\!\!\!
E\equiv \frac{1}{2} \left(\frac{dK}{dx}\right)^2+U\left(\rho[K(x)]\right)=\frac{{\tilde F}_4^2}{2\rho^5} \left(\frac{d\rho}{dx}\right)^2+U\left[\rho(x)\right],
\nonumber
\end{eqnarray}
is conserved. This can be verified directly from the one-dimensional version of Eq. (\ref{eq}),
\begin{eqnarray}&&\!\!\!\!\!\!
\frac{d}{dx}\left[\frac{{\tilde F}_4(\rho)}{\rho^{5/2}}\frac{d\rho}{dx} \right]=
f(\rho).
\end{eqnarray}
The resulting solution for $\rho(x)$ is implicitly given by 
\begin{eqnarray}&&
x=\int \frac{{\tilde F}_4(\rho) d\rho}{\rho^{5/2}\sqrt{2[E-U(\rho)]}},
\end{eqnarray}
where the constant of integration and $E$ should be determined using the boundary conditions. 
The physical significance of solutions depending on one coordinate only is obtained by considering the evolution of the
fluid in a long channel. Provided the channel is sufficiently narrow in the transverse direction it will remain macroscopically
uniform in those directions due to the stabilizing action of the heat conduction that is dominant at small scales (the same mechanism makes the uniform cooling state stable for small systems). Thus for long channels the fluid mechanical fields depend on $t$ and $x$ only. The solutions
to Eq.~(\ref{eq}) are natural candidates for the steady state of the system.

To perform the study of the IS and their role in the evolution of the system, it is necessary to have explicit expressions for the coefficients of the fluid mechanics of the fluid. The coefficients are known in the dilute limit, while in the case $\rho \sigma^3\sim 1$, one has to use certain approximations. Since the understanding is lacking in the case of the dilute gas already, then below we study if the IS provide the long-time state of the system in the case of the dilute granular gas in the long channel. Later we use this result to demonstrate the stability of the IS in the dense case too. 

In the case of the dilute gas one can write down the 
IS and the fluid mechanical equations explicitly. While the IS's form is known in the case of the gas \cite{IF,MFV}, its stability is only known for channels of moderate length $L\ll l_{cr}/\sqrt{1-r}$ that correspond to the fast sound limit, see \cite{MFV} and below. Here $\pi l_{cr}$ is the instability length that separates large systems where the uniform cooling state is unstable from the small ones where it is stable. In this work we settle positively the question whether the IS present the state of the gas at large times when $L$ is large but not necessarily bounded by $l_{cr}/\sqrt{1-r}$ . 

In the limit of small density we have ${\tilde F}_4\approx \kappa_0/(\gamma-1)$, where $\kappa_0$ is the thermal conductivity of the dilute gas,  ${\tilde F}\approx \Lambda \rho^{1/2}/(\gamma-1)$, where $\Lambda=\Lambda(\rho=0)$.  We find                      
\begin{eqnarray}&& \!\!\!\!\!\!\!\!\!\!\!\!
U=-\frac{\kappa_0\Lambda}{(\gamma-1)^2\rho}+\frac{2\kappa_0\Lambda\langle \rho^{1/2}\rangle}{3(\gamma-1)^2\rho^{3/2}},\label{potdil}\\&& \!\!\!\!\!\!\!\!\!\!\!\!
K(\rho)=\frac{2\kappa_0}{3(\gamma-1)\rho^{3/2}},  \ \ \rho^{1/2}=\left(\frac{2\kappa_0}{3K(\gamma-1)}\right)^{1/3}.\nonumber
\end{eqnarray}
We find that Eq. (\ref{isaac}) becomes  
\begin{eqnarray}&& \!\!\!\!\!\!
\frac{2\kappa_0}{3(\gamma-1)} \frac{d^2}{dx^2}\frac{1}{\rho^{3/2}} =\frac{\rho^{5/2}\partial_{\rho} U}{{\tilde F}_4(\rho)}=\frac{\Lambda\left[\rho^{1/2}-\langle \rho^{1/2}\rangle\right]}{(\gamma-1)}.
\nonumber\end{eqnarray}
The solution to this equation was found in \cite{MFV}, where the consideration relied on the use of the mass coordinate frame. In the next Section we describe the qualitative structure of the solution in the real space which will serve the basis for the study of the dense case. The quantitative description that reproduces the results of \cite{MFV} together with further details relevant to this work is performed in Section \ref{s7}. 

\section{Qualitative structure of the IS in one-dimensional case} 
\label{quality}

In this section we describe qualitatively the IS in the dilute one-dimensional case. Note that the study is quite similar to the consideration of soliton solutions in non-linear physics.
We use that
\begin{eqnarray}&& \!\!\!\!\!\!
\frac{d^2 K}{dx^2}=-\frac{\partial U}{\partial K},\ \ U=\left(\frac{2}{3}\right)^{1/3}
\frac{\Lambda \kappa_0^{1/3} }{(\gamma-1)^{4/3}}\biggl[K\langle K^{-1/3}\rangle 
\nonumber\\&& -\frac{3K^{2/3}}{2} \biggr],
\nonumber\end{eqnarray}
is identical in form to the Newton law of motion where $x$ is "time" and $K(x)$ is the "coordinate". 
We consider the behavior of $U(K)$ in the physically allowed domain of variation of $K$ which is $K\geq 0$. The potential has a minimum at $K_0=\langle K^{-1/3}\rangle^{-3}$ which is negative, $U(K_0)<0$. We have
\begin{eqnarray}&& \!\!\!\!\!\!
U\approx -\left(\frac{2}{3}\right)^{1/3}
\frac{\Lambda \kappa_0^{1/3} }{(\gamma-1)^{4/3}}\frac{3K^{2/3}}{2},\ \ 0<K\ll K_0;\nonumber\\&& 
U\approx \left(\frac{2}{3}\right)^{1/3}
\frac{\Lambda \kappa_0^{1/3} }{(\gamma-1)^{4/3}}K\langle K^{-1/3}\rangle,\ \ K_0\ll K.
\nonumber\end{eqnarray}
The uniformly cooling solution corresponds to the particle fixed in the minimum of the potential $K(x)\equiv K_0=\langle K\rangle$.  
The IS result from considering the finite periodic motion of the particle between the smaller and larger solutions $K_1(E)$ and $K_2(E)$ of the equation $E=U\left[K\right]$ where $U(K_0)<E<0$,
\begin{eqnarray}&& \!\!\!\!\!\!
x=\int_K^{K_2(E)} \frac{dK'}{\sqrt{2(E-U(K'))}},\label{solve}
\nonumber\end{eqnarray}
where the origin is chosen so that the minimum of the density $\rho$, that corresponds to the maximum $K_2(E)$ of $K$, is at $x=0$.
When $E$ is slightly larger than $U(K_0)$ the motion is harmonic, so that its period is finite. When $E$ increases, the period of the motion increases becoming infinite when $E$ tends to zero from below. The period becomes infinite because $K_0\propto \langle \rho^{1/2}\rangle^{-3}\propto L^{3/2}$ diverges in the thermodynamic limit, see the computation in the next sections. Since $K_2(E=0)=27 K_0/8$, then the period's divergence occurs due to the square root divergence of the integral in Eq. (\ref{solve}) at large $K$.   

The physical solution is determined from the condition that half the period of the periodic motion is equal to the size of the channel $L$,
\begin{eqnarray}&& \!\!\!\!\!\!
L=\int_{K_1(E)}^{K_2(E)} \frac{dK'}{\sqrt{2(E-U(K'))}}.
\nonumber\end{eqnarray}
Here we consider the no heat flux boundary conditions (b. c.) within which $\rho'(x)$ vanishes at the ends of the channel, implying that $K'(x)$ is also zero there (thus the ends of the channel correspond to the turning points in the solution). 
The solutions obtained by considering the full period of the motion (or any integer number of half-periods) obey the b. c. too, but they are not stable, see \cite{MFV} and below. 

Thus the IS exists for the channel length in the interval $L_{min}<L<\infty$, where $L_{min}$ corresponds to the finite period of the harmonic motion near the minimum of the oscillator ($L_{min}=\pi l_{cr}$, see \cite{MFV} and the next section). The density grows monotonously from its minimum at $x=0$ to the maximum at $x=L$. In the thermodynamic limit the minimum tends to zero (corresponding to divergence of $K_2[E=0]$), while the maximum tends to infinity (corresponding to $K_1[E=0]=0$). In particular, the dilute gas assumption breaks down in the thermodynamic limit. The resulting changes in the solution and omitted details are provided in the following sections.

\section{IS equations in the dilute gas limit}

In the limit of small density the fluid mechanical equations take a simpler form, where the viscosity coefficients and the thermal conductivity become proportional to $\sqrt{T}$, while $\Lambda$ becomes a constant. Using Eq.~(\ref{const}), we find
\begin{eqnarray}&&\!\!\!\!\!\!
\frac{\partial\rho}{\partial t}+\nabla\cdot[\rho \bm v]=0,\ \ \rho\left[\frac{\partial v_i}{\partial t}+(\bm v\cdot\nabla) v_i
\right]=-\nabla_i p+
\nonumber\\&&\!\!\!\!\!\!
+\nu\frac{\partial}{\partial x_k}\left[
\sqrt{\frac{p}{\rho}}
\left(\frac{\partial v_i}{\partial x_k}+\frac{\partial v_k}{\partial x_i}-\frac{2}{d}\delta_{ik}\nabla\cdot \bm v\right)\right],\nonumber\\&&
\!\!\!\!\!\!\left(\frac{\partial}{\partial t}\!+\!\bm v\cdot \nabla \right)p\!=\!
\!-\!\gamma p\nabla\cdot \bm v\!-\!\Lambda\rho^{1/2}p^{3/2}\!+\!\frac{2\kappa_0}{3}\nabla^2 \left(\frac{p}{\rho}\right)^{3/2}
\nonumber\\&&\!\!\!\!\!\!
+(\gamma-1)\nu
\sqrt{\frac{p}{\rho}}
\left[\frac{\partial v_i}{\partial x_k}\frac{\partial v_i}{\partial x_k}+\frac{\partial v_k}{\partial x_i}\frac{\partial v_i}{\partial x_k}-\frac{2}{d}\left(\nabla\cdot \bm v\right)^2\right],\label{fl}
\end{eqnarray}
where $\Lambda=2 \pi^{(d-1)/2} (1-r^2) \sigma^{d-1}/[d\,
\Gamma(d/2)]$ (see \textit{e.g.} \cite{Brey}), $\Gamma(\dots)$ is the
gamma-function, $\nu=(2\sigma\sqrt{\pi})^{-1}$ and $\kappa_0=4\nu$ in $d=2$, and
$\nu=5(4\sigma^2 \sqrt{\pi})^{-1}$ and $\kappa_0=5\nu/2$ in $d=3$ \cite{BP}. 

Equations~(\ref{fl}) differ from the
fluid mechanics of a dilute gas of \textit{elastically} colliding
spheres only by the presence of the inelastic loss term $-\Lambda \rho^{1/2} p^{3/2}$
which is proportional to the average energy loss per collision, $\sim (1-r^2)p$,
and to the collision rate, $\sim \rho^{1/2} p^{1/2}$ (remind that $p=\rho T$, where $T$ is the temperature).
As discussed above, the inequality $1-r\ll 1$ guarantees that the characteristic cooling time $1/\Lambda\rho^{1/2}p^{1/2}$ inferred from the equation on pressure is much larger than the mean free time $1/(\sigma^{d-1} \rho^{1/2}p^{1/2})$.

As already mentioned, the system of Eqs.~(\ref{fl}) has well-known homogeneous cooling solutions \cite{Haff}. These solutions
are described by Eqs.~(\ref{uniformly cooling state}) with the cooling time $t_c$ provided by the reduced expression
\begin{eqnarray}&&
t_c\equiv \frac{2}{\Lambda\rho_0^{1/2} p^{1/2}(0)}.
\end{eqnarray}
The solutions are known to be unstable with respect to the sinusoidal perturbations which wave-vector $k$ is smaller than $1/l_{cr}$
where the critical length $l_{cr}$ is given by 
\begin{eqnarray}&&
l_{cr}=\sqrt{\frac{2\kappa_0}{\Lambda\rho_0^2}},
\end{eqnarray}
see \cite{Goldhirsch1,Goldhirsch3,McNamara1,McNamara2}. Such wave-vectors exist when the system size exceeds $\pi l_{cr}$ for no-flux b. c. and when the system size exceeds $2\pi l_{cr}$ for p. b. c. Small perturbations with large wavelength
get enhanced and particles start to form clusters (the rotational modes that behave differently are not relevant for the present work). Correspondingly this instability is often called the clustering instability.

The length $l_{cr}$ also has a special physical meaning with respect to the IS, see the previous Section.
For system size below $\pi l_{cr}$ all the IS are homogenous, so the uniformly cooling states are the only states of the macroscopic rest of the system. In contrast, for larger system
size, while the uniformly cooling state continues to hold, inhomogeneous solutions appear as well \cite{MFV}. For no-flux b. c. the inhomogeneous solutions appear at the system size greater than $\pi l_{cr}$, while for the p. b. c. they appear at the system size greater than $2\pi l_{cr}$. This "coincidence" makes it natural to suggest that at these system sizes the inhomogeneous solutions become the attractors for the system's evolution in time, instead of the uniform cooling states holding for subcritical systems. This was proved in the limit of fast
sound \cite{MFV}. 

Finally, we present the equation on the density in the IS in the dilute limit. 
Setting $F_4=\kappa_0$ in Eq.~(\ref{eq}) one finds
\begin{eqnarray}&&\!\!\!\!\!\!
\rho_0^{1/2}-\langle \rho_0^{1/2}\rangle-\frac{2\kappa_0}{3\Lambda}\nabla^2\rho_0^{-3/2}=0.\label{eq1}
\end{eqnarray}
while the pressure is given by Eq.~(\ref{pres}) with $1/t_c=\Lambda\langle \rho_0^{1/2}\rangle p^{1/2}(0)/2$.
This equation and the numerical inhomogeneous solution for the spherically symmetric case were presented in \cite{IF}. 
The  Cauchy-Schwarz inequality implies $\langle \rho_0^{1/2}\rangle\leq \langle \rho_0\rangle^{1/2}$, so the energy decay 
for the inhomogeneous solutions is always slower than the one of the uniformly cooling state, see Eq.~(\ref{uniformly cooling state}).

The complete description of the solutions to the non-linear PDE (\ref{eq1}) is likely to be available only numerically.
Thus even in the limit of the dilute gas of inelastic hard spheres, neither the complete description of the IS, not the understanding of their relevance
to the evolution of the system are available. The case that allows progress is the case of fields depending on only one spatial coordinate, to the study of which we pass. Though the solutions that we describe below were obtained previously \cite{MFV,IF}, their consideration is necessary here. This is because the thermodynamic limit was never considered in detail and because these solutions are needed to consider the dense IS.

\section{IS depending on one coordinate for dilute granular gas}
\label{s7}

We now concentrate on the study of the solutions to Eq.~(\ref{eq1}) that depend only on the coordinate $x$ and obey 
\begin{eqnarray}&&\!\!\!\!\!\!
\frac{2\kappa_0}{3\Lambda}\frac{d^2}{dx^2}\rho_0^{-3/2}=\rho_0^{1/2}-\langle \rho_0^{1/2}\rangle.\label{eq2}
\end{eqnarray}
This equation is relevant for long channels with length $L$, so it is considered in the interval $(0, L)$. 
We consider two kinds of boundary conditions, where the solutions are slightly different - the periodic boundary condition (p. b. c.) and the no heat flux boundary condition. For the considered solution with the spatially uniform pressure the condition of no heat flux gives the condition of vanishing derivative of $\rho_0$ at the boundary. It is convenient to measure the density in the units of average density $\rho_0$ and the distance in the units of $l_{cr}$. We find that the rescaled density $\rho'$ obeys in the new coordinate $x'$ the equation
\begin{eqnarray}&&
\rho'^{1/2}-\langle \rho'^{1/2}\rangle'-\frac{1}{3}\frac{d^2}{dx^2}\rho'^{-3/2}=0,\nonumber\\&&
\langle \rho'\rangle'\equiv \frac{1}{{\cal L}}\int_0^{\cal L}\rho'(x')dx'=1,\label{eq3}
\end{eqnarray}
where the last condition follows from $\langle \rho'\rangle=1$ and uniformity of $\rho'$ in the transversal directions.  
The rescaled length ${\cal L}=L/l_{cr}$ of the channel is
\begin{equation}
{\cal L}=\frac{L}{l_{cr}}=
\left\{\begin{array}{ll}
 (\sqrt{\pi}/2)(1-r^2)^{1/2} \rho_0 \sigma L\;\;\;\quad \mbox{in 2d,} \\
\sqrt{16\pi/75}\,(1-r^2)^{1/2} \rho_0 \sigma^2 L \quad \mbox{in 3d}\,.
\end{array}
\right.
\label{calL}
\end{equation}
Equation (\ref{eq3}) and the solution to it were obtained in \cite{MFV}. Here we reproduce the solution with the purpose of discussing its thermodynamic limit. It is convenient to pass to the mass coordinate
\begin{eqnarray}&&\!\!\!\!\!\!\!\!\!\!\!
m(x')\equiv \int_0^{x'}\rho(x'')dx'',\ \ x'(m)=\int_0^m \frac{dm'}{\rho'(m')},\label{eqt1}
\end{eqnarray}
where the solution's interval of definition is the same interval $(0, {\cal L})$ since in the rescaled variables, the rescaled length of the channel ${\cal L}$ coincides with the rescaled total mass of the gas, $\int_0^{{\cal L}}\rho'(x')\,dx'$, cf. Eq.~(\ref{eq3}). The condition (\ref{eq3}) that the average density is one, is substituted in the mass coordinate frame by the condition that the "average length" equals one:
\begin{eqnarray}&&
\frac{1}{{\cal L}}\int_0^{\cal L}  \frac{dm}{\rho'(m)}=\int_0^{\cal L} \frac{dx}{{\cal L}}=1.\label{eq4}
\end{eqnarray}
We obtain the following equation for $w\equiv \rho'^{-1/2}$
\begin{eqnarray}&&
\frac{d^2 w}{dm^2}=w-w^2\langle w\rangle_m, \label{basic}
\end{eqnarray}
where the angular brackets with the subscript designate the "spatial" average over $m$.
Equation~(\ref{basic}) is defined on the interval $0<m<{\cal L}$, at the ends of which we demand zero first derivative of $w$, which corresponds to the no-flux boundary conditions. To get rid of the (\textit{a priori} unknown) factor $\langle w \rangle_m$, we introduce a new variable
\begin{eqnarray}&&
f(m)=\left\langle w\right\rangle_m w(m)
\end{eqnarray}
and obtain\begin{equation}\label{ordinary}
\frac{d^2 f}{dm^2}=f-f^2\,.
\end{equation}
Integrating the above equation from $0$ to ${\cal L}$, we find that both for p. b. c. and no-flux b. c., the averages of $f$ and $f^2$ coincide. Since $\langle f\rangle_m=\left\langle w\right\rangle_m^2$ and $\langle f^2\rangle_m=\left\langle w\right\rangle_m^2\langle w^2\rangle_m$ we conclude that the condition of conservation of "average length" $\langle w^2\rangle_m=1$ is obeyed automatically, once the b. c. are imposed on  $f$.
After $f$ is found, one can restore $w$ via
\begin{equation}
\label{old} w=\frac{f}{\sqrt{\left\langle f\right\rangle_m}}.
\end{equation}
Equation~(\ref{ordinary}) has appeared in numerous applications, and its solutions are well known.  We consider $f$ as a coordinate of a Newtonian particle of unit mass, moving in the potential $U(f)=f^3/3-f^2/2$. The
``total energy" $E$ is conserved:
\begin{eqnarray}&&
E=\frac{1}{2}\left(\frac{df}{dm}\right)^2+\frac{f^3}{3}-\frac{f^2}{2}.\label{energyintegral}
\end{eqnarray}
The boundary conditions can be obeyed only by bounded solutions with $-1/6\leq E\leq 0$, where we can write
\begin{equation}
\frac{f^3}{3}-\frac{f^2}{2}-E=\frac{(f-a[E])(f-b[E])(f-c[E])}{3}, \label{roots}
\end{equation}
where $a[E]>b[E]>c[E]$ are the real roots of the cubic polynomial that give the turning points of the trajectory where the velocity vanishes. 
Here we stressed that these roots are functions of the "energy" $E$. 
The no-flux b. c. condition demands that the "initial coordinate" $f(0)$ and the "final coordinate" $f({\cal L})$ are either $a$ or $b$ (since $c<0$ these are the only physically meaningful turning points). The solutions obeying the p. b. c. can be obtained by gluing together the solutions with no-flux b. c., so that $f(0)=f({\cal L})$ is either $a$ or $b$. Thus we first consider the solutions obeying the no-flux b. c.
A bounded solution of Eq.~(\ref{ordinary}) can be written as
\begin{eqnarray}&&
m(f)=\int_{f}^{a(E)}\frac{df'}{\sqrt{2E-2f'^{3}/3+f'^2}}. \label{solution}
\end{eqnarray}
This solution obeys $m[a(E)]=0$ and $m'[a(E)]=\infty$ so the above solution satisfies the correct boundary condition at $m=0$,
\begin{eqnarray}&&
\frac{df}{dm}|_{m=0}=0.
\end{eqnarray}
This solution is constructed so that $f(m)$ reaches its maximal value $a(E)$ at $m=0$. The first positive zero $m_1$ of $f'(m)$ determined by (\ref{solution}) is given by
\begin{eqnarray}&&
m_1=\int_{b(E)}^{a(E)}\frac{df'}{\sqrt{2E-2f'^{3}/3+f'^2}}. \label{solution0}
\end{eqnarray}
In particular if we consider the "fundamental" solution with no zeros of $f'(m)$ at $0<m<{\cal L}$, then the "energy"
$E({\cal L})$ of the solution corresponding to length ${\cal L}$ is determined from
\begin{eqnarray}&&\!\!\!\!\!\!\!\!\!\!\!\!\!\!
{\cal L}=\int_{b[E({\cal L})]}^{a[E({\cal L})]}\frac{df}{\sqrt{2E({\cal L})-2f^{3}/3+f^2}}\label{solution1}\\&&\!\!\!\!\!\!\!\!\!\!\!\!\!\!
=\sqrt{\frac{6}{a[E({\cal L})]-c[E({\cal L})]}}{\bm K}\left(\sqrt{\frac{a[E({\cal L})]-b[E({\cal L})]}{a[E({\cal L})]-c[E({\cal L})]}}\right), \label{resolution}
\end{eqnarray}
where the value of the integral and the definition of the complete elliptic integral ${\bm K}(x)$ can be found in \cite{Gradshteyn}. 
The choice of the initial condition made above corresponds to $f(m)$ that monotonously decreases from $f(0)=a[E({\cal L})]$ to
$f({\cal L})=b[E({\cal L})]$. The solution ${\tilde f}(m)$ for which $f(m)$ monotonously grows from $f(0)=b[E({\cal L})]$ to
$f({\cal L})=a[E({\cal L})]$ can be obtained as $f({\cal L}-m)$, which gives
\begin{eqnarray}&&
{\cal L}-m({\tilde f})=\int_{{\tilde f}}^{a[E({\cal L})]}\frac{df'}{\sqrt{2E({\cal L})-2f'^{3}/3+f'^2}}. \label{solution2}
\end{eqnarray}
Using Eq.~(\ref{solution1}) we may also write
\begin{eqnarray}&&
m({\tilde f})={\cal L}-\int_{{\tilde f}}^{a[E({\cal L})]}\frac{df'}{\sqrt{2E({\cal L})-2f'^{3}/3+f'^2}}\nonumber\\&&
=\int_{b[E({\cal L})]}^{{\tilde f}}\frac{df'}{\sqrt{2E({\cal L})-2f'^{3}/3+f'^2}}.
\end{eqnarray}
The above solution, of course, could also be obtained directly. The usefulness of this solution is that it being glued with the previous solution 
it gives the fundamental solution for the p. b. c. This solution is also as relevant for the evolution toward the IS as the previous solution. 

Returning to Eq.~(\ref{solution}), using the formula from p. $234$ of \cite{Gradshteyn} and the
definition of the elliptic integral, we find
\begin{eqnarray}&&
\sqrt{\frac{a-c}{6}}m=\int_0^{\arcsin \sqrt{(a-f)/(a-b)}}\frac{d\alpha}{\sqrt{1-\frac{(a-b)\sin^2\alpha}{a-c}}}. \nonumber
\end{eqnarray}
Next, using the definitions from p. $924$ of \cite{Gradshteyn}, we find
\begin{eqnarray}&&
\!\!\!\!\!\!\!\!\!\!f(m)=c+(a-c)\,{\mbox dn}^2\left(\sqrt{\frac{a-c}{6}}m,\, \sqrt{\frac{a-b}{a-c}}\right)\,,
\label{solve}
\end{eqnarray}
where ${\mbox dn}$ is one of the Jacobi elliptic functions. To write down the solution for $w(m)$ we use the value of the integral from p. $644$ of \cite{Gradshteyn},
\begin{eqnarray}&&\!\!\!\!\!\!\!\!\!\!\!\!
\langle f\rangle_m=c[E({\cal L})]+(a[E({\cal L})]-c[E({\cal L})])\frac{{\bm E}\left(\sqrt{s[E({\cal L})]}\right)}{{\bm K}\left(\sqrt{s[E({\cal L})]}\right)}\nonumber\\&&\!\!\!\!\!\!\!\!\!\!\!\!
\equiv C^2({\cal L}), \ \ \ \ s[E({\cal L})]\equiv \frac{a[E({\cal L})]-b[E({\cal L})]}{a[E({\cal L})]-c[E({\cal L})]}
\label{defconst}
\end{eqnarray}
where ${\bm E}(x)$ is the complete elliptic integral of the second kind. The function $C({\cal L})$ has a very important role for the IS because it determines the decay rate of the pressure for these solutions. We have 
\begin{eqnarray}&&\!\!\!\!\!\!
\langle \rho_0^{1/2}\rangle={\bar \rho}^{1/2}\frac{1}{{\cal L}}\int_0^{{\cal L}} \rho'^{1/2}(x')dx'=\rho_0^{1/2}\langle w\rangle_m
\nonumber\\&&
=\rho_0^{1/2}\langle f\rangle_m^{1/2}=\rho_0^{1/2}C({\cal L}).
\end{eqnarray}
It follows that the pressure for the IS is given by 
\begin{eqnarray}&&
p(t)=\frac{p(0)}{\left[1+t/t_c\right]^2},\ \ t_c\equiv 
\frac{2}{C({\cal L})\Lambda \rho_0^{1/2} p^{1/2}(0)}, \nonumber
\end{eqnarray}
Thus $C({\cal L})$ determines the deviations of the decay time from the decay time of the uniformly cooling state, and as we saw 
one must have $C({\cal L})\leq 1$ with equality holding only for uniformly cooling state. Finally, using Eqs.~ (\ref{old}) and (\ref{solution}), we write the 
solution for $\rho'(m)$:
\begin{equation}\label{w(m)}
    \frac{1}{\rho'(m)^{1/2}}
    =\frac{c+(a-c)\, {\mbox dn}^2\left(\sqrt{\frac{a-c}{6}}\,m, \sqrt{\frac{a-b}{a-c}}\right)}{C({\cal L})}\,,
\end{equation}
We now pass to consider solutions derivable from the fundamental solution above, and the solutions' limits for different system size. 

\subsection{Periodic boundary conditions and solutions with multiple reflections}

We constructed above the solution that is monotonic in $(0, {\cal L})$. We called this solution "fundamental" as the rest of the solutions can be obtained from it by gluing it with the reflected solution. If we reflect the above solution and glue it with $w(2{\cal L}-m)$ we get the fundamental periodic solution for the system with length $2{\cal L}$. Further application of reflections and gluing produce solutions with multiple reflections at the turning points. Numerical simulations indicate the the solution with the minimal possible number of the turning points is the one which is stable, cf. \cite{MFV}. Thus for no-flux boundary conditions the solution that is stable is the fundamental solution described above, while for the p. b. c. the stable solution is the fundamental periodic solution described above. 
 
\subsection{The critical lower length for the existence of the IS}

Clearly for any system size ${\cal L}$ there are solutions with $f=1$ where the particle stands indefinitely at the potential minimum at $f=1$. This is the uniformly cooling state described above. The inhomogeneous solutions correspond to the deviation of the particle from the minimum of the potential and these solutions have a minimal period corresponding to the harmonic expansion of the potential near the minimum. The existence of this minimal period signifies that inhomogeneous solutions exist only for ${\cal L}$ larger than a certain critical length. This length is fixed by considering 
$E=-1/6+\delta E$, $0<\delta E \ll 1$. In this limit, the effective "Newtonian" particle is a harmonic oscillator with $U(f)\approx -1/6+(f-1)^2/2$. It follows that for no-flux b. c. the fundamental solution is $f(m)=1+\sqrt{2\delta E} \cos m$ and $w(m)=1+\sqrt{2\delta E} \cos m$, where we noticed $\langle f\rangle_m=1$. These solutions exist only above the critical length ${\cal L}=\pi$ and are a small-amplitude sinusoidal modulation of the uniformly cooling state $w(m)=1$. For the p. b. c. the solution has the same form and it exists above the critical length ${\cal L}=2\pi$. 
The expressions for $E({\cal L})$ can be obtained by considering the usual corrections to the independence of the period of the amplitude.

The sinusoidal solutions for slightly supercritical systems described above were checked numerically to provide the universal state of the gas after long time of evolution. 
The description of the numerical results is provided later.

\subsection{The IS in the thermodynamic limit}

Our main interest here is the solution for large ${\cal L}$. At ${\cal L}\gg 1$ the
correspondence between the energy $E$ and length ${\cal L}$ is $|E|\approx 72 \exp[-2{\cal L}]$.
This can be found by noting that at small $|E|$ we have $a\approx 3/2$, $b\approx {\sqrt{2|E|}}$ and $c\approx -{\sqrt{2|E|}}$. 
Using that at $z$ close to unity
\begin{eqnarray}&&
K(z)=-\frac{1}{2}\ln(1-z^2)+\ln 4+\ldots,
\end{eqnarray}
where $\ldots$ vanish at $z=1$, we find
\begin{eqnarray}&&
{\bm K}\left(\sqrt{\frac{a-b}{a-c}}\right)\approx -\frac{1}{2}\ln\left(\frac{b-c}{a-c}\right)\nonumber\\&&+\ln 4\approx -\frac{1}{4}\ln |E|+\frac{1}{4}\ln 72.
\end{eqnarray}
It follows from Eq.~(\ref{resolution}) that the relation between $E$ and ${\cal L}$ at large system size is $|E|=72\exp[-2{\cal L}]$. Note the
difference of the factor of $2$ from \cite{MFV}: it arises due to the use of no-flux, rather than periodic, boundary conditions, see above.

Thus the thermodynamic limit of ${\cal L}\to\infty$ corresponds to $|E|\to 0$. To study this limit
we consider the solution
\begin{eqnarray}&&
m(f)=\int_{f}^{a(E)}\frac{df'}{\sqrt{2E({\cal L})-2f'^{3}/3+f'^2}}. \label{solutionanalysis1}
\end{eqnarray}
at $|E|\to 0$. In the lowest order approximation we set $E=0$ above which gives
\begin{eqnarray}&&\!\!\!\!\!\!\!\!\!\!
m(f)\!=\!\int_{f}^{3/2}\frac{df'}{f'\sqrt{1-2f'/3}}\!=\!\ln\left(\frac{1+\sqrt{1-2f/3}}{1-\sqrt{1-2f/3}}\right).\label{solutionanalysis0}
\end{eqnarray}
Inverting the above relation we obtain
\begin{eqnarray}&&
f=\frac{3}{2\cosh^2\left(m/2\right)},\ \ {\cal L}-m\gg 1. \label{solutionanalysis2}
\end{eqnarray}
where the condition follows from negligibility of the term $E$ in the denominator of Eq.~(\ref{solutionanalysis1}). Due to ${\cal L}\gg 1$ the above asymptotic form covers almost all the interval $(0, {\cal L})$, however there is a vicinity of $m={\cal L}$ that is not described by Eq.~(\ref{solutionanalysis2}). This approximation used to derive Eq.~(\ref{solutionanalysis2}) becomes invalid as $f$ approaches zero ($f$ reaches $b$ which is small), which is signalled by the divergence of $m(f)$
in Eq.~(\ref{solutionanalysis0}) at $f=0$. To study the vicinity of $f=0$ we write
\begin{eqnarray}&&
m(f)={\cal L}-\int_{b(E)}^{f}\frac{df'}{\sqrt{2E({\cal L})-2f'^{3}/3+f'^2}}. \label{solutionanalysis3}
\end{eqnarray}
At $|E|\to 0$ we have $b(E)\approx \sqrt{2|E|}\to 0$, so considering $b(E)\leq f\ll 1$ we have
\begin{eqnarray}&&
{\cal L}-m(f)\approx \int_{\sqrt{2|E|}}^{f}
\frac{df'}{\sqrt{f'^2-2|E({\cal L})|}}
\nonumber\\&&=
\cosh^{-1}
\frac{f}{\sqrt{2|E|}}. \label{solutionanalysis4}
\end{eqnarray}
Using $\sqrt{2|E|}\approx 12 \exp[-{\cal L}]$ we find
\begin{eqnarray}&&
f\approx 12 e^{-{\cal L}}\cosh({\cal L}-m),\ \ m\gg 1.
\end{eqnarray}
where the condition $m\gg 1$ corresponds to  $f\ll 1$. It is immediate from the expressions above that $\langle f\rangle$ is determined by $m\ll {\cal L}$ where one can use Eq.~(\ref{solutionanalysis2}),
\begin{eqnarray}&&
C^2({\cal L})=\langle f\rangle\approx \frac{3}{{\cal L}},
\end{eqnarray}
that can also be obtained directly by expanding Eq.~(\ref{defconst}) at small $|E|$. We find that in the limit ${\cal L}\gg 1$, the pressure 
obeys
\begin{eqnarray}&&
p(t)=\frac{p(0)}{\left[1+t/t_c\right]^2},\ \ t_c\equiv 
\frac{2\sqrt{L}}{\sqrt{3l_{cr}}\Lambda\rho_0^{1/2} p^{1/2}(0)}. \nonumber
\end{eqnarray}
We observe that the decay time grows with the system size in contrast to the uniformly cooling state having decay time which is local. This behavior can be traced to the power-law $\rho(x)\propto (L-x)^{-1}$ that holds in a wide region of scales. This law would diverge in the non-integrable way at $x=L$. Thus the integral for the mass $\int \rho(x) dx$ is determined by the right cut-off of $(L-x)^{-1}$ near $x=L$. In contrast 
$\int \rho^{1/2} dx$, that determines the decay time, converges at $x=L$, so that it is determined by the left cut-off of the $(L-x)^{-1}$ law producing $\int \rho^{1/2}dx \sim L^{1/2}$ that corresponds to 
$\langle \rho^{1/2}\rangle\propto L^{-1/2} $. 

Thus the IS are solutions where the system acts as a single whole so that local measurements would be affected by the global structure of the solution. For $w$ we find 
\begin{eqnarray}&&
w\approx \frac{\sqrt{3{\cal L}}}{2}\frac{1}{\cosh^2\left(m/2\right)},\ \ {\cal L}-m\gg 1
,\nonumber\\&& \ w\approx \sqrt{48{\cal L}} e^{-{\cal L}}\cosh({\cal L}-m),\ \ m\gg 1.
\end{eqnarray}
The two of the above asymptotic expressions overlap in the wide interval and together they cover $(0, {\cal L})$ completely. For the density $\rho'=w^{-2}$ we have
\begin{eqnarray}&&
\rho'\approx \frac{4}{3{\cal L}}\cosh^4\left(m/2\right),\ \ {\cal L}-m\gg 1
,\label{solutionanalysis7}\\&& \ \rho'\approx \frac{e^{2{\cal L}}
}{48{\cal L}\cosh^2({\cal L}-m)},\ \ m\gg 1.\label{solutionanalysis8}
\end{eqnarray}
In the joint asymptotic region the density grows exponentially according to $\rho'\approx \exp[2m]/12{\cal L}$. 

In the thermodynamic limit ${\cal L}\gg 1$ the solution for $\rho'$ is determined by Eq.~(\ref{solutionanalysis7}). Indeed, the mass coordinate 
gives directly the mass of the gas described by the corresponding asymptotic expression. The condition ${\cal L}-m\gg 1$ signifies that 
in the thermodynamic limit a fraction $y$ of the mass of the gas that is arbitrarily close to unity, but such that ${\cal L}(1-y)\gg 1$ is described 
by the asymptotic expression (\ref{solutionanalysis7}).
To write down the corresponding
expressions in real space we use the inverse transformation $x'(m)$ from Eq.~(\ref{eqt1}).  
We find
\begin{eqnarray}&&
\!\!\!\!\!\!\!\!x'=\frac{3{\cal L}}{2}\int_0^{m/2}\frac{dm'}{\cosh^4 m'}=\frac{3{\cal L}}{2}\left[\tanh\left(\frac{m}{2}\right)-\frac{\tanh^3 \left(\frac{m}{2}\right)}{3}\right].\nonumber
\end{eqnarray}
Thus $w(x')$ is determined implicitly by
\begin{eqnarray}&&
\!\!\!\!\!\!\!\!\frac{2x'}{3{\cal L}}=\tanh\left(\frac{m}{2}\right)-\frac{\tanh^3 \left(\frac{m}{2}\right)}{3}
, \label{solutionanalysis5}\\&&
\frac{2w}{\sqrt{3{\cal L}}}\approx 1-\tanh^2\left(\frac{m}{2}\right).
\end{eqnarray}
This formula can be made explicit using the substitution $\tanh (m/2)=2\sin\phi$. This solves explicitly the cubic polynomial in the first line changing it into $x'/{\cal L}=\sin 3\phi$, 
so that the distribution of mass in space obeys
\begin{eqnarray}&&
\tanh \left[\frac{m(x')}{2}\right]=2\sin\left(\frac{\arcsin(x/{\cal L})}{3}
\right)
\end{eqnarray}
provided ${\cal L}-m(x')\gg 1$. We obtain 
\begin{eqnarray}&&\!\!\!\!\!\!\!\!\!\!\!\!\!\!
m(x')\approx \frac{4x'}{3{\cal L}},\ \ x\ll {\cal L},\nonumber \\&& \!\!\!\!\!\!\!\!\!\!\!\!\!\!
m(x')\!\approx \!\frac{1}{2}\ln\left(\frac{6{\cal L}}{{\cal L}-x'}\right),\ \ \!  \exp[-2{\cal L}] \ll 1\!-\!\frac{x'}{{\cal L}}\ll 1,
\label{mass}
\end{eqnarray}
where we noted that the condition ${\cal L}-m(x')\gg 1$ gives ${\cal L}+\ln[1-x'/{\cal L}]/2\gg 1$ or $1-x'/{\cal L}\gg \exp[-2{\cal L}]$.
It follows that the mass of the system concentrates in the neighbourhood of ${\cal L}$, defined by $1-x'/{\cal L}\ll 1$ (the mass in the region $(0, {\cal L}[1-\epsilon])$ is proportional to $\ln 1/\epsilon$ 
which is negligible in comparison with the total mass ${\cal L}$ if ${\cal L}\gg \ln 1/\epsilon$
, cf. below). 
Noting that $w/\sqrt{3{\cal L}}+1/2=\cos2\phi$, we obtain in physical variables
\begin{eqnarray}&&
\rho(x)=\frac{4\rho_0 l_{cr}}{3L}\left[2\cos\left[\frac{2}{3}\arcsin\left(\frac{x}{L}\right)\right]-1\right]^{-2},\nonumber\\&& 
1-\frac{x}{L}\gg \exp\left[-\frac{2L}{l_{cr}}\right], \label{solutionanalysis6}
\end{eqnarray}
where the condition is equivalent to ${\cal L}-m(x')\gg 1$. This formula is equivalent to the formula provided in \cite{MFV}. Note that, as explained above, for large enough ${\cal L}$ one can extend the above expression to such $x'$ that the mass fraction $m(x')/{\cal L}$ is close to unity and the expression describes the distribution of the major part of the mass of the gas. In fact, Eq.~(\ref{solutionanalysis6}) covers almost all gas length $(0, {\cal L})$ excluding exponentially small vicinity of $x={\cal L}$. Performing the expansion at $x/L\ll 1$ or $x/L\approx 1$, to directly differentiating $m(x')$ one obtains 
\begin{eqnarray}&&
\rho(x)\approx \frac{4\rho_0 l_{cr}}{3L},\ \ x\ll L/l_{cr},\nonumber \\&& 
\rho(x)\approx \frac{\rho_0 l_{cr}}{2\left(L-x\right)},\ \ 
 \exp\left[-\frac{2 L}{l_{cr}}\right] \ll1-\frac{x}{L}\ll 1.\nonumber
\end{eqnarray}
It is illuminating to write the results in the form
\begin{eqnarray}&&\!\!\!\!\!\!\!
\rho(x)\!=\!\frac{4\rho_0 l_{cr}}{3L}\left[2\cos\left[\frac{2}{3}\arcsin\left(\frac{x}{L}\right)\right]\!-\!1\right]^{-2},\ \ \rho\ll \rho_{max},\nonumber\\&&\!\!\!\!\!\!\!
\rho(x)\approx \frac{\rho_0 l_{cr}}{2\left(L-x\right)},\ \ \frac{\rho_0 l_{cr}}{L} \ll \rho(x)\ll \rho_{max}.\label{das}
\end{eqnarray}
This form shows clearly the behavior of the density. The density field has large variation in space, changing from its value $4\rho_0 l_{cr}/3L$ in the dilute phase at $x\ll L$ to 
$\rho \sim \rho_{max}$ in the vicinity of $x=L$. The interpolation between the two regions follows a power law $[L-x]^{-1}$.  

The first of the equations above shows that the density has self-similar scaling in $L$: one has $\rho(x)={\tilde F}(x/ L)/L$. It is non-obvious how this form can describe mass that grows linearly with $L$ (we keep $\rho_0$ constant) since $\int_0^{{\cal L}}{\tilde F}(x/L)dx/L$ would give an ${\cal L}-$independent quantity. The resolution to this apparent paradox is that Eq.~(\ref{solutionanalysis6}) applies roughly up to $x=L-L\exp[-2{\cal L}]$ and the integral would diverge at $x=L$,
\begin{eqnarray}&&
\int_0^{{\cal L}-{\cal L}\exp[-2{\cal L}]}\frac{4dx'}{3{\cal L}}\left[2\cos\left[\frac{2}{3}\arcsin\left(\frac{x'}{{\cal L}}\right)\right]-1\right]^{-2}
\nonumber\\&&
\sim \int_0^{{\cal L}-{\cal L}\exp[-2{\cal L}]} \frac{dx'}{2{\cal L}-2x'}\sim \frac{1}{2}\ln \frac{2{\cal L}}
{2{\cal L}\exp[-2{\cal L}]}\sim {\cal L}.\nonumber
\end{eqnarray}
We now show that in fact the domain $\rho\ll \rho_{max}$ contains the larger part of the system's mass, while the neighborhood of the maximum defined by 
$\rho\sim \rho_{max}$ contains mass of order one. This can be seen from Eq.~(\ref{solutionanalysis8}) that shows that the density decays away from the maximum exponentially. Thus $\rho'(m)\ll \rho'_{max}$ when ${\cal L}-m\gg 1$. Say, $\rho'\left[m={\cal L}-3\right]\approx 4\exp[-6]\rho_{max}\ll \rho_{max}$. Since the difference of the mass coordinates measures the mass in physical space, then we conclude that the mass contained in the region
$\rho\sim \rho_{max}$ is of order one. This mass is much smaller than the total "mass" ${\cal L}$ in the considered limit. 

Finally, to describe the whole interval $(0, {\cal L})$ we use Eq.~(\ref{solutionanalysis8}) employing the relation between $x'$ and $m$ in the form
\begin{eqnarray}&&
{\cal L}-x'=\int_{m}^{{\cal L}}\frac{dm'}{\rho(m')}.
\end{eqnarray}
Confining the above expression to $m\gg 1$ we may use Eq.~(\ref{solutionanalysis8}) to find
\begin{eqnarray}&&
{\cal L}-x'=48{\cal L}e^{-2{\cal L}}\left[\frac{{\cal L}-m}{2}+\frac{\sinh[2({\cal L}-m)]}{4}
\right].
\end{eqnarray}
The above expression together with Eq.~(\ref{solutionanalysis8}) determine implicitly the profile of the density in the region not covered
by the previous asymptotic expressions. In the region $m\gg 1$ and ${\cal L}-m\gg 1$ the above
equation reproduces the power-law behavior of the density. 
In the region ${\cal L}-m\ll 1$, not captured by the previous results, we find
\begin{eqnarray}&&
{\cal L}-x=48{\cal L}e^{-2{\cal L}}\left({\cal L}-m\right).
\end{eqnarray}
Using the above equation and Eq.~(\ref{solutionanalysis8}) we find
 \begin{eqnarray}&&
\rho\approx \frac{\rho_{max}}{\cosh^2\left[\rho_{max}({\cal L}-x)\right]},\ \ \rho_{max}\equiv  \frac{e^{2{\cal L}}
}{48{\cal L}},\nonumber\\&& \ \rho_{max}({\cal L}-x)\ll 1.\nonumber
\end{eqnarray}
Keeping above the $\cosh({\cal L}-m)$ term, and not expanding it at ${\cal L}-m\ll 1$ is a matter of convenience. The above form makes it
obvious that the density has a maximum which width is inverse to the maximum.

To summarize, in the thermodynamic limit one can use Eq.~(\ref{solutionanalysis6}) in the major part of the system. This expression however would diverge at $x={\cal L}$ in a non-integrable way, and it needs to be cut off at the maximal density $\rho_{max}$. The latter grows exponentially with the system size, though the mass contained in the region $\rho\sim \rho_{max}$ is of order one. 

The indefinite growth of $\rho_{max}$ with the system size shows that the consistent consideration of the thermodynamic limit can not be made within the frame of the dilute gas approximation even 
if the condition $\rho_0\sigma^3\ll 1$ is satisfied. The consideration demands studying the IS of the dense fluids introduced in the previous sections and their stability. Nevertheless, the first step to understanding the stability of the IS is to study that in the dilute gas approximation. Then, as described in the Introduction, this can be used to derive the stability of the dense IS. Thus we pass to the analysis of the question whether the IS constitute the final state of the fluid in the dilute gas approximation.

\section{The IS as the universal long-time limit of evolution}

It was shown in \cite{MFV} that the IS
is the attractor for the long-time evolution of the gas in the limit of fast sound. Within this limit the sound travel time through the system $t_s\sim L/\sqrt{T}$ is assumed to be much smaller than the characteristic time-scale of the cooling $t_c\sim 1/\Lambda \rho_0\sqrt{T}$, so that
$L\ll 1/\Lambda \rho_0$. Since $1/\Lambda \rho_0 \sim l_{cr}/\sqrt{1-r^2}$, then the fast sound limit is the case $L\ll l_{cr}/\sqrt{1-r^2}$. Thus the limit of the fast sound allows non-trivial values of $L\geq \pi l_{cr}$ only for $1/\sqrt{1-r^2}\ll 1$ which is a more restrictive inequality than 
$1-r^2\ll 1$ needed for the validity of the hydrodynamic approach as such. Due to the assumption $t_s\ll t_c$, the pressure becomes uniform throughout the gas faster than any effects due to inelasticity take place. Thus the latter effects can be analyzed assuming they develop on the background of a uniform pressure. Clearly this limit does not allow
to address the thermodynamic limit $L\to \infty$.

Thus we study the system behavior in the thermodynamic limit, which is probably the most important physical question about the considered system.
As we argued, for long channels the macroscopic fields depend on only one spatial coordinate $x$. The evolution of these fields is then described by the corresponding reduction of the system (\ref{fl}) that reads
\begin{eqnarray}&&
\frac{\partial \rho}{\partial t}+\frac{\partial(\rho v)}{\partial x}=0,
\label{fl1}\\&& \rho\left(\frac{\partial v}{\partial t}+
v\frac{\partial v}{\partial x}\right)=-\frac{\partial p}{\partial
x}+\nu_0\frac{\partial}{\partial x}\left(\sqrt{\frac{p}{\rho}}\frac{\partial v}{\partial
x}\right), \label{fl2}
\\&& \frac{\partial p}{\partial t} +v \frac{\partial p}{\partial x}=-\gamma
p\frac{\partial v}{\partial x}-\Lambda\rho^{1/2} p^{3/2}\nonumber\\&&
+\kappa_0 \frac{\partial}{\partial x}
\left[\sqrt{\frac{p}{\rho}}\,\frac{\partial}{\partial x}
\left(\frac{p}{\rho}\right)\right]+\nu_0(\gamma-1)\sqrt{\frac{p}{\rho}}
\left(\frac{\partial v}{\partial x}\right)^2. \label{fl3}
\end{eqnarray}
where $\nu_0=4\nu/3$ in $d=3$ and $\nu_0=\nu$ in $d=2$. The system is considered for $0<x<{\cal L}$ where ${\cal L}\equiv L/l_{cr}$ is the channel length $L$ measured in the units of $l_{cr}$. The system should be supplied with the appropriate boundary condition. We will assume rigid, insulating walls when both particles and heat flux vanish, $v(x=0)=v(x=L)=0$ and $\partial_x T(x=0)=\partial_x T(x=L)=0$, where the ideal gas relation $T=p/\rho$ should be used.

The above system needs to be solved at the average value of the three-dimensional density equal to one. Since the
density is uniform in transversal directions, then the average one-dimensional density is also one
\begin{eqnarray}&&
\frac{1}{{\cal L}}\int_0^{{\cal L}}\rho(x) dx = 1.
\end{eqnarray}
This equation is a constraint on the solutions. The thermodynamic limit corresponds to considering the limit ${\cal L}\to \infty$
at average one-dimensional density fixed at one.
We study if at large times the solutions to the above system tend to the IS 
\begin{eqnarray}&&\!\!\!\!\!\!\!\!\!
\rho=\rho_0\left(\frac{x}{l_{cr}}\right),\ \ v=0,\ \ 
p(t)=\frac{p(0)}{\left[1+C({\cal L})t/t_c^0\right]^2},
\end{eqnarray}
where $t_c^0$ is the decay time of the uniformly cooling states and $\rho_0(x)$ is the IS's density profile. 
The convergence to the IS would signify that for the supercritical systems the density profile saturates at large times at an inhomogeneous profile,
\begin{eqnarray}&&
\lim_{t\to\infty}\rho(x, t)=\rho_0(x).
\end{eqnarray}
For the pressure we would like to check the existence of the following limit
\begin{eqnarray}&&
\lim_{t\to\infty} p(x, t)[1+C({\cal L})t/t_c^0]^2=p_0,
\end{eqnarray}
with some effective constant $p_0$. There is no need to check separately the corresponding convergence of the velocity as it is implied by the relations above.

Below we measure distances in the units of $l_{cr}$ and times in the units of $t_c^0$ where
instead of $p(0)$ one uses $p_0$. Thus we assume that the IS is the attractor for the system evolution, so there is a certain value of $p_0$, 
and we check the self-consistency of this assumption. We also pass to dimensionless fields and measure
density in the units of $\rho_0$, velocity in the units of $l_{cr}/t_c^0$ and pressure in the units of $p_0$. Keeping with no ambiguity the original notation for the fields and the coordinates we find that the following dimensionless form of the system (\ref{fl1})-(\ref{fl3}) holds in $d=3$,
\begin{eqnarray}&&
\frac{\partial\rho}{\partial t}+\frac{\partial (\rho v)}{\partial x}=0,\ \
\nonumber\\&&
\varepsilon_1\rho\left[
\frac{\partial v}{\partial t}+v\frac{\partial v}{\partial x}
\right]=-\frac{\partial p}{\partial x}
+\varepsilon_2 \frac{\partial}{\partial x}\left[
\sqrt{\frac{p}{\rho}}
\left(\frac{\partial v}{\partial x}\right)\right],\nonumber\\&&
\frac{\partial p}{\partial t}+v\frac{\partial p}{\partial x}=
-\gamma p\frac{\partial v}{\partial x}
-2\rho^{1/2}p^{3/2}+\frac{2}{3}\frac{\partial^2}{\partial x^2} \left(\frac{p}{\rho}\right)^{3/2}
\nonumber\\&&
+\varepsilon_2(\gamma-1)
\sqrt{\frac{p}{\rho}}
\left(\frac{\partial v}{\partial x}\right)^2.\label{hydrodynamiceqoo}
\end{eqnarray}
where $\varepsilon_1=\kappa_0\Lambda/2$ and $\varepsilon_2=2\nu\Lambda/3$. Note that $\varepsilon_1\sim \varepsilon_2\sim 1-r^2\ll 1$.

It is more convenient to study the solution $\rho_0$ and its attracting properties by using the mass coordinate frame. This is defined by the passage from coordinates $[x, t]$ to $[m(x, t), t]$ where
\begin{eqnarray}&&
m(x, t)=\int_0^x \rho(x', t)dx',\ \ \partial_t m+v\partial_x m=0,
\end{eqnarray}
where the last equation uses that the gas velocity vanishes at the boundary $v(x=0, t)\equiv 0$ [of course also $v(x={\cal L}, t)\equiv 0$]. It follows from the above that the inverse transformation $x(m, t)$ is a Lagrangian coordinate,
\begin{eqnarray}&&
\frac{\partial x(m, t)}{\partial t}=v[x(m, t), t],
\end{eqnarray}
which means simply that the end point of the interval $[0, x(m, t)]$ containing a given mass $m$ moves with the fluid.
The equations take a somewhat simpler form in $[m, t]$ coordinates:
\begin{eqnarray}&&
\frac{\partial}{\partial t}\frac{1}{\rho}=\frac{\partial v}{\partial m},
\label{hydrodynamics1}\\&& \varepsilon_1\frac{\partial v}{\partial t}=-\frac{\partial p}{\partial
m}+\varepsilon_2\frac{\partial}{\partial m}\left(\sqrt{p\rho}\frac{\partial v}{\partial
m}\right), \label{hydrodynamics2}
\\&& \frac{\partial p}{\partial t}=-\gamma
p \rho\frac{\partial v}{\partial m}-2\rho^{1/2} p^{3/2}
+ \rho \frac{\partial}{\partial m}
\left[\sqrt{p\rho}\,\frac{\partial}{\partial m}
\left(\frac{p}{\rho}\right)\right]. \nonumber\\&&
+\varepsilon_2 \,(\gamma-1)\, \rho^{3/2}p^{1/2}
\left(\frac{\partial v}{\partial m}\right)^2. \label{hydrodynamics3}
\end{eqnarray}
By a transformation similar to the one described in Sec. \ref{transformation} we pass to the variables in which the IS solution is time-independent. We introduce fields $p'$ and $v'$ by 
\begin{eqnarray}&&
p=\frac{p'}{[1+C({\cal L})t]^2},\ \ v=\frac{v'}{1+C({\cal L})t},
\end{eqnarray}
and the new time variable 
\begin{eqnarray}&&
\tau=\frac{1}{C({\cal L})}\ln [1+C({\cal L})t],\ \ \frac{d\tau}{dt}=\frac{1}{1+C({\cal L})t}.
\end{eqnarray}
In the new field and variables the system takes the form 
\begin{eqnarray}&&
\frac{\partial}{\partial \tau}\frac{1}{\rho}=\frac{\partial v'}{\partial m},
\nonumber\\&& \varepsilon_1\frac{\partial v'}{\partial \tau}-\varepsilon_1 C({\cal L}) v'=
-\frac{\partial p'}{\partial
m}+\varepsilon_2\frac{\partial}{\partial m}\left(\sqrt{p'\rho}\frac{\partial v'}{\partial
m}\right), \nonumber
\\&& \frac{\partial p'}{\partial \tau}-2C({\cal L}) p'=-\gamma
p' \rho\frac{\partial v'}{\partial m}-2\rho^{1/2} p'^{3/2}\nonumber\\&&
+ \rho \frac{\partial}{\partial m}
\left[\sqrt{p'\rho}\,\frac{\partial}{\partial m}
\left(\frac{p'}{\rho}\right)\right]. \label{rescaledIS}
\end{eqnarray}
The IS solution in these variables has a very simple form:
\begin{eqnarray}&&
\rho=\rho_0(m), \ \ p'=1,\ \ v'=0.
\end{eqnarray}
These variables are significantly more convenient for numerical studies than the original variables for which the IS is time-dependent. 
We have performed the numerical studies of the system of Eqs.~(\ref{rescaledIS}). The studies of the thermodynamic limit ${\cal L}\to\infty$ 
appear impossible due to the exponential growth of the maximal density with the system size. We have succeeded in performing simulations up to the system size ${\cal L}=8$. For this size the maximal density is about $23100$. We have observed that the IS is the global attractor of the system 
dynamics at large times. Further increase in the system size appears impractical within the frame of the direct numerical simulations. Say, 
for system size ${\cal L}=9$ the maximal density is already about $152000$. Clearly a special device is needed to study the system's relaxation to the IS for the decade of ${\cal L}\gg 1$. 

The simulations were performed for the no-flux boundary conditions. We used the value of $\gamma=2$ of the two-dimensional gas and the values of 
$\varepsilon_1=1-r^2$ and $\varepsilon_2=(1-r^2)/4$ for $r=0.98$. This value of $r$ does not give a large value of $1/\sqrt{1-r^2}$ and consequently there is no non-trivial region of applicability of the fast sound regime in this case. Thus no theoretical prediction on the relevance of the IS exists in this case. The simulations showed that for the supercritical systems with ${\cal L}>\pi$, 
the IS are stable attractors for ${\cal L}\leq 8$. While the uniformly cooling state is the steady state of the system at ${\cal L}<\pi$, at  $\pi<{\cal L}\leq 8$, the place of the uniformly cooling state is taken by the IS. These states are both linearly and non-linearly stable, that is they are the universal attractors of the system evolution in time for arbitrary initial conditions. We pass to the description of the results of the numerical simulations. 


\label{s8}

\section{Results of the numerical simulations}

\begin{figure}
\includegraphics[width=6.0 cm,clip=]{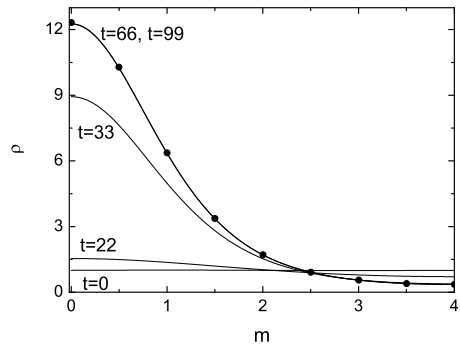}
\caption{The evolution of the density to $\rho_0$ for ${\cal L}=4$. The plot of $\rho_0$ is marked by circles} \label{LongPaperRoOfML4}
\end{figure}

\begin{figure}
\includegraphics[width=6.0 cm,clip=]{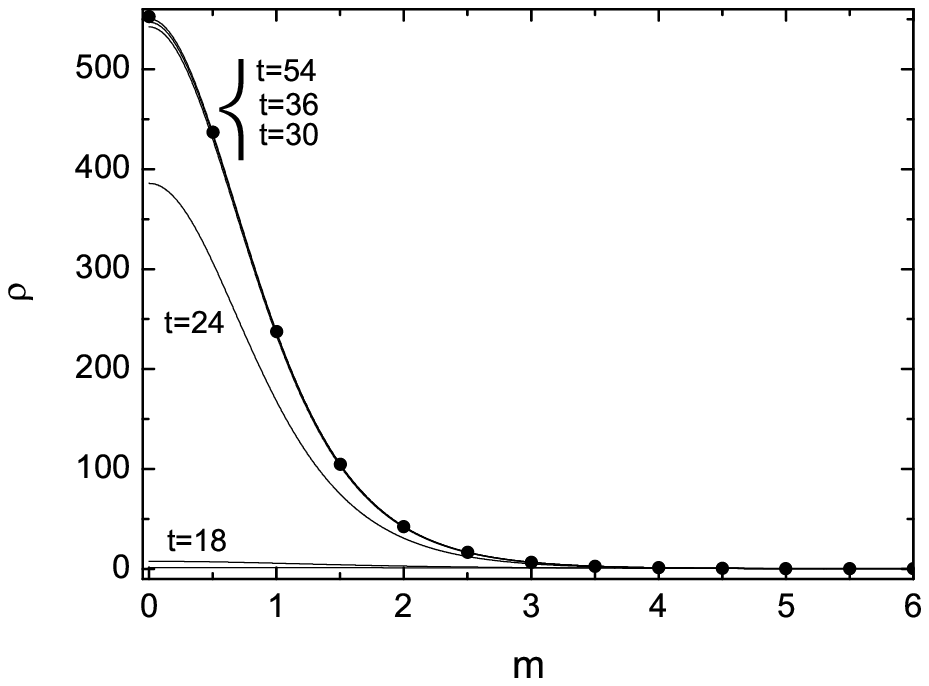}
\caption{The evolution of the density to $\rho_0$ for ${\cal L}=6$. The plot of $\rho_0$ is marked by circles} \label{LongPaperRoOfML6}
\end{figure}

\begin{figure}
\includegraphics[width=6.0 cm,clip=]{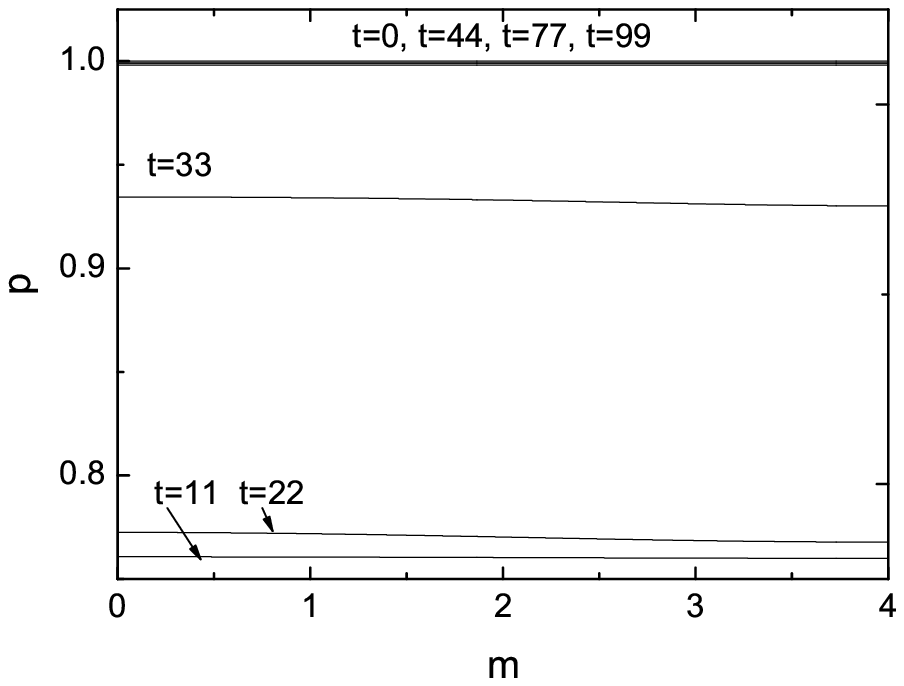}
\caption{The evolution of the pressure to unity for ${\cal L}=4$} \label{LongPaperPofML4}
\end{figure}

\begin{figure}
\includegraphics[width=6.0 cm,clip=]{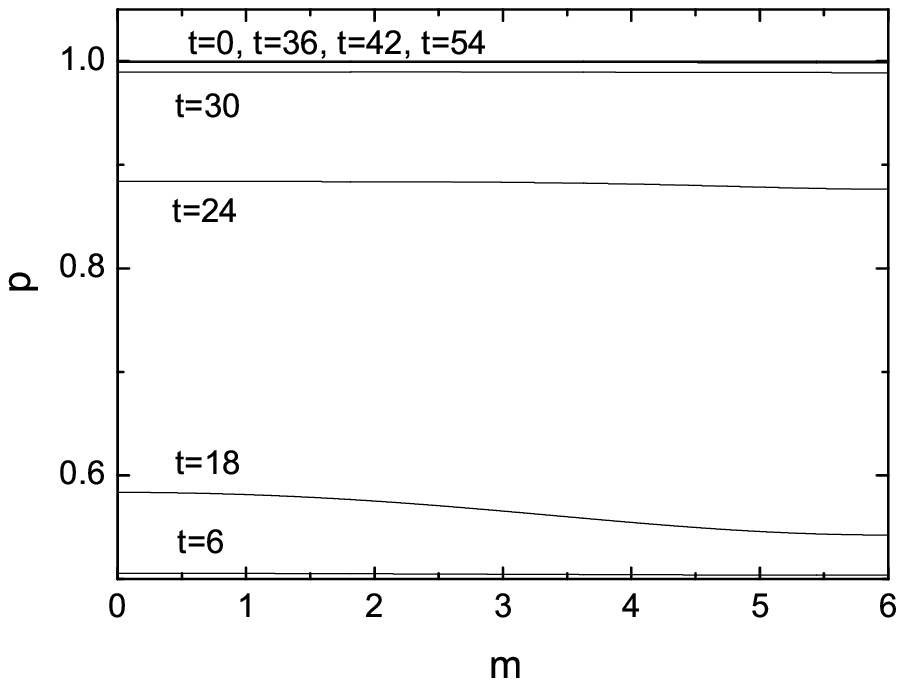}
\caption{The evolution of the pressure to unity for ${\cal L}=6$} \label{LongPaperPofML6}
\end{figure}

The results of the simulations for the system's size ${\cal L}=4$, ${\cal L}=6$, ${\cal L}=7$ and ${\cal L}=8$ are 
shown in Figs.~(\ref{LongPaperRoOfML4})-(\ref{LongPaperPofML6}). The evolution clearly brings the initial conditions to the IS.
The relaxation is exponential.  In this section we use $t$ instead of $\tau$, so in physical time the relaxation is a power law. 
\begin{figure}
\includegraphics[width=7.0 cm,clip=]{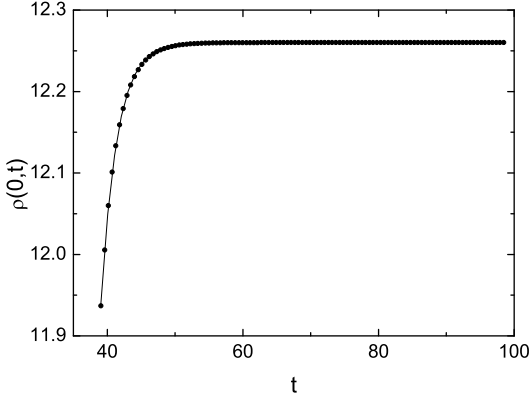}
\caption{The evolution of the maximal density to the steady state value for ${\cal L}=4$. The best fit is
$\rho_{max}=12.3-3925814\exp(-t/2.39)$.} \label{Length4MaxRho=12.3-3925814exp(-t2.39)}
\end{figure}
\begin{figure}
\includegraphics[width=7.0 cm,clip=]{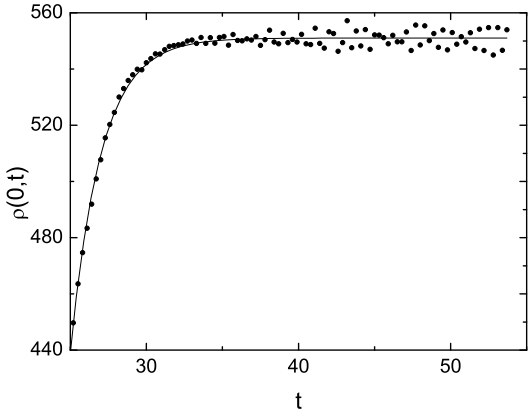}
\caption{The evolution of the maximal density to the steady state value for ${\cal L}=6$. The best fit is
$\rho_{max}=551-27000000\exp(-t/2.02)$.} \label{Length6MaxRho=551-27000000exp(-t/2.02)}
\end{figure}
\begin{figure}
\includegraphics[width=7.0 cm,clip=]{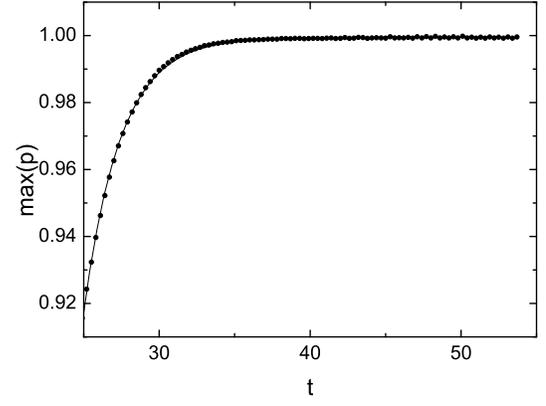}
\caption{The evolution of the maximal pressure to unity for ${\cal L}=6$. The best fit is
$p_{max}=0.99953-2298\exp(-t/2.44)$.} \label{Length6MaxP=0.99953-2298exp(-t2.44)}
\end{figure}
\begin{figure}
\includegraphics[width=8.0 cm,clip=]{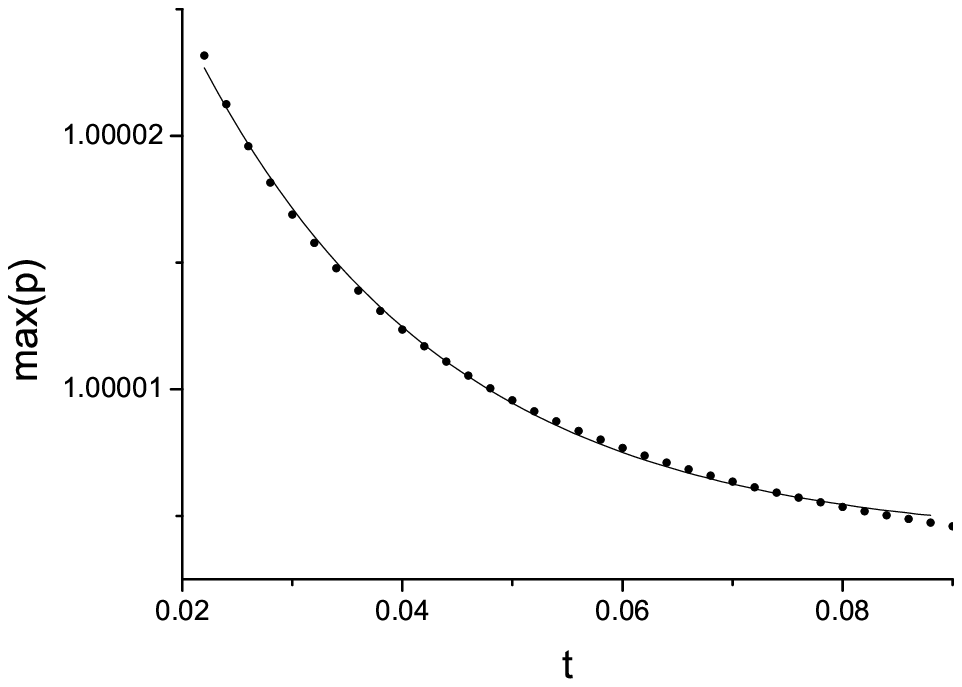}
\caption{The long-time exponential relaxation of the maximal pressure to unity for ${\cal L}=7$. The best fit is
$p_{max}=1+0.00005\exp(-t/0.0227)$.} \label{Length7MaxP=-1+0point00005exp(-t/0point0227)}
\end{figure}
\begin{figure}
\includegraphics[width=8.0 cm,clip=]{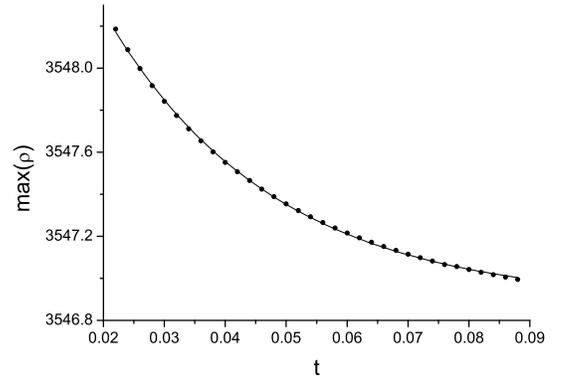}
\caption{The long-time exponential relaxation of the maximal density to the steady state value for ${\cal L}=7$. The best fit is 
$\rho_{max}=3546+2.8565\exp(-t/0.0276)$.} \label{Length7MaxRho=3546+2point8565exp(-t0point0276)}
\end{figure}
\begin{figure}
\includegraphics[width=8.0 cm,clip=]{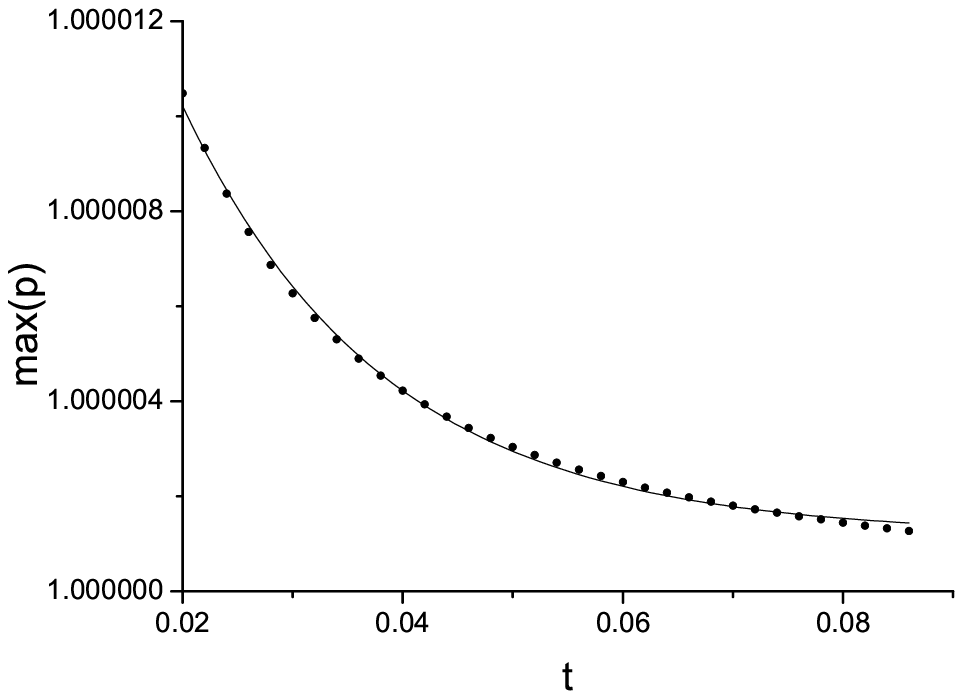}
\caption{The long-time exponential relaxation of the maximal pressure to unity for ${\cal L}=8$. The best fit is
$p_{max}=1+0.00003\exp(-t/0.0183)$.} \label{Length8MaxP=1+0.00003exp(-t0.0183)}
\end{figure}
\begin{figure}
\includegraphics[width=8.0 cm,clip=]{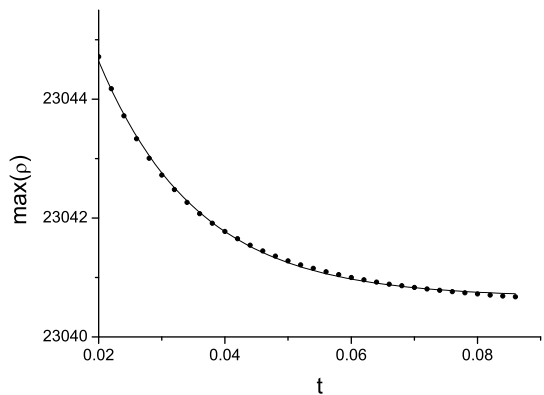}
\caption{The long-time exponential relaxation of the maximal density to the steady state value for ${\cal L}=8$. The best fit is
$\rho_{max}=23040+14.3\exp(-t/0.0156)$.} \label{Length8MaxRho=23040+14.3exp(-t0.0156)}
\end{figure}

The numerical simulations allow to discuss besides the basic fact of the relaxation of the initial conditions to the IS (which if exists is exponential by the equations), also the dependence of the relaxation exponent on ${\cal L}$. We studied the exponents of relaxation of the density and the pressure. The two exponents must correspond to the same eigenmode of the linearized operator described in the previous Section and hence to be equal. This is confirmed by the results of the numerical experiments where the exponents coincide with the numerical accuracy. 

The dependence of the exponent $\lambda$ that describes the exponential relaxation to the IS at large times exhibits remarkably strong dependence on ${\cal L}$. The exponent does not change much from ${\cal L}=4$ (where $\lambda\sim 0.42$) to ${\cal L}=6$ (where $\lambda\sim 0.42$). However, at 
${\cal L}=7$ the exponent jumps to $\lambda\sim 40$ and at ${\cal L}=8$ one has $\lambda\sim 60$. These numbers are given with the accuracy of $10$ to $20$ per cent and they are not an artifact of a numerical problem: the decay fit to the exponential relaxation is extremely good as it is clear from the Figures.  Thus the decay exponent $\lambda$ experiences a significant jump by about a factor of $10$ as one passes from ${\cal L}=6$ to ${\cal L}=7$. This jump apparently signifies that between those values of ${\cal L}$ the system enters the asymptotic region of large sizes ${\cal L}\gg 1$ with the corresponding change of the functional dependence of $\lambda$ on ${\cal L}$. The study of $\lambda({\cal L})$ at large ${\cal L}$ is beyond the current abilities of the numerical experiment and the explanation of the jump is postponed for future work. Here we would only like to establish the fundamental fact that the IS is the attractor of the system's evolution at large times.

\label{numerics}

\section{The IS in the dense case}
\label{dense}

In this section we consider the IS in the dense case without presuming that $\rho\sigma^3\ll 1$ holds everywhere. This is always the case in the thermodynamic limit where the maximal density grows exponentially with the system size. The properties of the IS of the dilute gas that were derived in the previous sections show that the dilute gas assumption breaks down when the system size $L$ obeys $\rho_{max}=\rho_0 l_{cr}\exp[2L/l_{cr}]/48 L\sim \rho_p$ where $\rho_p\equiv\sigma^{-3}$ is of the order of the density of close packing. In this case in the region of maximal density the diluteness breaks down. We describe the resulting changes in the distributions of mass and energy throughout the system. 

It is assumed that the total number of particles $N$ obeys $N\sigma^3/\Omega=\rho_0\sigma^3 \ll 1$ so that on average the fluid is still dilute (which is, in particular, the case of the clustering instability).  
Then the conservation of mass implies that the dense, liquid regions that form in the system occupy the volume's fraction that is much less than unity. It can be expected then that the IS that hold in the dilute case will be changed so that the dilute gas solution holds approximately in the interval $0\leq x\leq l_g$ while at $x\geq l_g$ the IS is different due to the density comparable with $\rho_p$. This is proved below. Note that $1-l_g/L\ll 1$. 

We use that the solution can be described by $K(x)$ that obeys 
\begin{eqnarray}&&
\frac{d^2K}{dx^2}=-\frac{\partial U(K)}{\partial K}, 
\end{eqnarray}
where $U(K)=U\left[\rho(K)\right]$ with 
\begin{eqnarray}&& \!\!\!\!\!\!\!\!\!\!\!\!\!\!\!\!\!\!\!\!
U(\rho)=\int \frac{{\tilde F}_4(\rho){\tilde F}(\rho)}{\rho^{5/2}}d\rho-
\frac{\langle  {\tilde F}(\rho)\rangle}{\langle \rho/F_1(\rho)\rangle} \int 
\frac{{\tilde F}_4(\rho) d\rho}{\rho^{3/2} F_1(\rho)},\label{definition}
\end{eqnarray}
where $\rho(K)$ is a monotonically decreasing function of $K$. The corrections due to the finite size of the particles change 
the dilute gas relation (\ref{potdil}) for $U(K)$ at small $K\sim K(\rho_p)$. Further change in $U(K)$ is caused by the changes in the averages $\langle  {\tilde F}(\rho)\rangle$ and 
$\langle \rho/F_1(\rho)\rangle$ due to the deviation of the IS from the dilute gas solution. We demonstrate that the change in the latter is negligible. We observe that the ratio of the pressure to the pressure of the dilute gas $p/\rho T$ is a growing function of $\rho$. Thus $\rho/F_1(\rho)=\rho T/p\leq 1$. It follows that the integral of $\rho/F_1(\rho)$ over the yet-to-be-found solution is the integral over a positive function that equals one in almost all the volume except the dense region of the liquid where this function is bounded between zero and one. We conclude that the contribution of the dense region in the integral can be neglected producing 
$\langle \rho/F_1(\rho)\rangle \approx 1$. Similarly one can demonstrate that $\langle  {\tilde F}(\rho)\rangle$ taken on the IS with dense regions is close to the one that would hold for the dilute gas. The reason is that in the case of the dilute gas where ${\tilde F}(\rho)\propto \rho^{1/2}$ the integral $\int \rho^{1/2}$ is determined by the dilute region far from the maximum (in which the liquid forms). This will be shown below using self-consistency. 

Thus we can confine the consideration of changes in $U(K)$ due to the formation of the liquid to the consideration of the role of the changes in the functional form of $U(K)$ at small $K\sim K(\rho_p)$. First we note that the functions that appear in the definition (\ref{definition}) of $U(\rho)$ remain bounded and smooth when $\rho$ approaches $\rho_p$.  Consider first 
the function ${\tilde F}_4(\rho)\equiv F_4(\rho)F_1'(\rho)\rho^{5/2}/F_1^{5/2}(\rho)$ that appears both in $U(\rho)$ and the definition of $K(\rho)$. The heat conductivity of the fluid of hard spheres $F_4$ stays finite in the limit of dense packed fluid. In contrast, the pressure described by $F_1=p/T$ grows indefinitely when the fluid gets denser. To see the impact of this divergence 
we consider the most divergent term in the Carnahan-Starling equation of state, 
\begin{eqnarray}&& 
F_1(\rho)/\rho=1+\rho F'(\rho)=\frac{1}{[1-\rho/\rho_p]^2},
\nonumber\end{eqnarray}
where the notation $\rho_p$ is taken for clarity, see the definitions (\ref{CS}), (\ref{tildeF}). 
We observe that though $F_1$ diverges at $\rho=\rho_p$, still ${\tilde F}_4$ is finite due to the division by $F_1^{5/2}$. This conclusion holds for all equations of state where the divergence at $\rho=\rho_p$ is described by the power-law term $F_1(\rho)\propto [1-\rho/\rho_{max}]^{-k}$ if $k>2/3$.  Since the latter condition is to hold for the 
true constitutive relation, then we assume that the conclusion that ${\tilde F}_4(\rho)$ is finite holds for the true constitutive relation. Finally ${\tilde F}$ described by Eq. (\ref{tildeF}) remains finite in the dense limit too. 

We conclude that the change in $U(K)$ caused by the finite-density is finite. This change is such that $K=0$ becomes a forbidden value of $K(x)$, so that $K(x)$ cannot go to un-physical values smaller than $K(\rho_p)$. 
Further we note that in the thermodynamic limit the energy $E$ will still approach $E=0$ since the divergence of the period of the particle's motion occurs in the region of large $K$ or small density, see Section \ref{quality} (this follows from the fact that the major part of the volume is occupied by the dilute gas so that the time that the particle spends at large $K$ is infinite in the thermodynamic limit). Thus $U(K)$ has to increase at $K\sim K(\rho_p)$ so that the smaller turning point at zero energy, defined by $U(K)=0$, is given by a finite $K$ of order $K(\rho_p)$. Correspondingly we assume that the change in $U(K)$ is such that it increases smoothly at $K\sim K(\rho_p)$ in comparison with the dilute gas value but remains monotonously decreasing. Note that the minimum of the potential is realized at $K_0$ that diverges in the thermodynamic limit and thus occurs at the density of the dilute gas, so the finite-density changes only the behavior of $U(K)$ at $K\ll K_0$, far from the potential's minimum. We assume that there is no new extremum of $U(K)$ at small $K$ which seems physically necessary. Thus $U(K)$ decreases monotonously from $U[K(\rho_p)]$ to its minimum at $K=K_0$ where $K_0$ belongs to the dilute gas region. 

It follows that the qualitative structure of the solution described in Section \ref{quality} is not changed by the finite particles' size.
These changes become relevant when the energy $E$ of the solution is such that the smallest positive solution to $E=U(K)$ is comparable with $K(\rho_p)$. They change the correspondence between $L$ and $E$: the "particle" with coordinate $K(x)$ spends different time at small $K$. Thus we introduce $E_{dense}(L)$ as the dependence of $E$ on $L$ determined by the condition that the half the period of the periodic motion with energy $E_{dense}(L)$ equals $L$. The solution is given by 
\begin{eqnarray}&& \!\!\!\!\!\!
x=\int_K^{K_2[E_{dense}(L)]} \frac{dK'}{\sqrt{2[E_{dense}(L)-U(K')]}}.
\nonumber\end{eqnarray}
It follows that in the region of the dilute gas where $U(K)$ is the same as in the dilute case, the solution is the same as in the dilute case. The only change is that instead of $E(L)$ that would correspond to the considered $L$ in the dilute gas limit $\sigma\to 0$, one has to use $E_{dense}(L)$. Introducing $L_{eff}(L)$ by $E\left[L_{eff}(L)\right]=E_{dense}(L)$ (this is possible because both $E(L)$ and $E_{dense}(L)$ are monotonic) we can say that the solution in the dilute region is like for the dilute gas with the effective length of the channel $L_{eff}(L)$.

Thus the solution in the dense case has the following structure. The density profile starts from its minimum at $x=0$ where the dilute gas holds. The density increases monotonously like it would for the dilute gas in the channel with the length $L_{eff}(L)$ (note that $L_{eff}(L)$ tends to infinity in the thermodynamic limit). Then at a certain scale $l_g$ the dilute gas assumption breaks down, so that in the region $(l_g, L)$ the density obeys $\rho\sigma^{-3}\sim 1$. Since there is no sharp boundary between the phases, then $l_g$ is defined up to a factor of order one, which will be seen inessential for the final result.  The total mass $m_g$ of the dilute gas in the region $(0, l_g)$ can be described by the formula (\ref{mass}) with $L_{eff}(L)$ instead of $L$ . The condition $m_g+m_l={\cal L}$ that the total mass equals ${\cal L}$ gives ($x_g=l_g/l_{cr}$) 
\begin{eqnarray}&&
\frac{1}{2}\ln\left(\frac{6{\cal L}_{eff}
}{{\cal L}_{eff}-x_g}\right)+\left({\cal L}-x_g\right)\left[c_l\rho_0\sigma^3\right]^{-1}={\cal L}
,\nonumber
\end{eqnarray}
where $m_l=\left({\cal L}-x_g\right)\left[c_l\rho_0\sigma^3\right]^{-1}$ is the mass of the liquid contained in the region $(l_g, L)$
with $c_l$ a constant of order one. Since $m_g$ depends on ${\cal L}$ only logarithmically and $\rho_0\sigma^3\ll 1$, then we find that at large ${\cal L}$ (one divides the equation by ${\cal L}$ and takes the limit ${\cal L}\to\infty$),
\begin{eqnarray}&&
1-\frac{l_g}{L}\approx c_l\rho_0\sigma^3,\ \ l_g\approx L(1-c_l\rho_0\sigma^3),\label{lg}
\end{eqnarray}
that is almost all the mass of the system is contained in the liquid phase. To determine $m_g$ we note that since $\rho(l_g)\sim \sigma^{-3}$ then we can use for $\rho(l_g)$ the asymptotic form in Eq. (\ref{das}). This 
gives the self-consistency condition 
\begin{eqnarray}&&
\rho_0(l_g)\approx \frac{\rho_0 l_{cr}}{2\left(L_{eff}-l_g\right)}\sim \sigma^{-3},\nonumber
\end{eqnarray}
which gives
\begin{eqnarray}&&
L_{eff}-l_g \sim \rho_0 \sigma^{3} l_{cr}.\nonumber
\end{eqnarray}
Since $l_{cr}\ll l_g$ then $L_{eff}\approx l_g$, that is the effective length of the channel is where the liquid phase starts. In other words, the beginning of the liquid phase is like a wall boundary condition for the gas. 
It follows that the mass of the gas phase is
\begin{eqnarray}&&
m_g\approx \frac{1}{2}\ln\left(\frac{6L
}{\rho_0 \sigma^{3}l_{cr}}\right),\nonumber
\end{eqnarray}
where the approximate equality holds with logarithmic accuracy. Thus the mass of the gas is infinite in the thermodynamic limit, however, it is only logarithmically large in the system size, demonstrating that the gas represent a vanishing fraction of the total mass of the system. 

The energy of the fluid is however determined by the gas phase and not by the liquid phase. Indeed, the energy density is given by $\rho T/(\gamma-1)$. Throughout the region of the dilute gas, which occupies most of the volume, this coincides with $p(t)/(\gamma-1)$. It follows that the energy density is uniform through most of the volume, deviating from the constant only in the liquid region. In the latter region the energy density can be written as $p(t)\rho/(\gamma-1)F_1(\rho)$, so that it is bounded from above by $p(t)/(\gamma-1)$. We find that the total energy $E(t)$ obeys
\begin{eqnarray}&&
E(t)=\frac{p(t)\Omega}{\gamma-1}\langle \rho/F_1(\rho) \rangle\approx \frac{p(t)\Omega}{\gamma-1}, 
\end{eqnarray}
where we used $\langle \rho/F_1(\rho) \rangle\approx 1$ derived previously. It follows that in the considered case where the gas is dilute on average, $\rho_0\sigma^3\ll 1$, so the gas phase volume is close to $\Omega$, we have that almost all the energy of the system is contained in the gaseous phase. 

The conclusion that though the liquid phase contains the fraction of the total mass that is close to unity, its energy is negligible, can be understood by noting that the temperature of the gas particles is much higher than of the liquid ones. Consider for example, $x\ll L$ where $\rho(x)\approx 4\rho l_{cr}/3L$, so that the temperature there $T(x)\approx 3Lp(t)/4\rho l_{cr}$ grows linearly with the size of the system. Thus though the gas particles are not many, their velocity is so high that they give dominant contribution into the system's energy.  

Finally, we consider the decay time of the solution $t_c$ 
\begin{eqnarray}&&
t_c\!\equiv\! 
\frac{2\langle \rho F_1^{-1}(\rho)\rangle}{\langle \Lambda(\rho)\rho^{3/2}F_1^{-1}(\rho)\rangle p^{1/2}(0)},
\end{eqnarray}
see Eq. (\ref{pres}). We observed previously that $\langle \rho F_1^{-1}(\rho)\rangle\approx 1$, so it remains to consider  $\langle \Lambda(\rho)\rho^{3/2}F_1^{-1}(\rho)\rangle=(\gamma-1)\langle {\tilde F}\rangle$. We saw previously that ${\tilde F}$ remains finite in the dense region, hence we can write 
\begin{eqnarray}&&
\int_0^L {\tilde F}dx\sim \int_0^{l_g} {\tilde F}dx+c_l\rho_0\sigma^3 L{\tilde F}(l_g),
\end{eqnarray}
where we used Eq. (\ref{lg}) and noted that continuity and finiteness of ${\tilde F}$ imply ${\tilde F}(l_g)\sim {\tilde F}(L)$ (the latter is because the density throughout the liquid phase preserves its order of magnitude $\rho_p$). The first integral can be found using ${\tilde F}(x)$ in the dilute phase, 
\begin{eqnarray}&& 
\int_0^{l_g} {\tilde F}dx \approx \frac{\Lambda}{\gamma-1} \int_{1}^{l_g}\frac{\rho_0^{1/2}l_{cr}^{1/2}dx}{2^{1/2}(l_g-x)^{1/2}},
\nonumber\end{eqnarray}
where one can write approximate equality because the integral is determined by $x\ll l_g$ that is $\langle \Lambda(\rho)\rho^{3/2}F_1^{-1}(\rho)\rangle$ is determined by $x$ inside the dilute phase far from the boundary of the liquid. In particular, this implies that $\langle \Lambda(\rho)\rho^{3/2}F_1^{-1}(\rho)\rangle$ is approximately the same as for the dilute IS with $L_{eff}$ instead of $L$. Using that $L_{eff}\approx L$ we conclude that $t_c$ coincides with the one of the dilute IS, 
\begin{eqnarray}&&
t_c\approx 
\frac{2}{C({\cal L})\Lambda \rho_0^{1/2} p^{1/2}(0)}
,\nonumber \\&& 
t_c\approx 
\frac{2\sqrt{L}}{\sqrt{3l_{cr}}\Lambda\rho_0^{1/2} p^{1/2}(0)},\ \ L\gg l_{cr}. \nonumber
\end{eqnarray}
Thus the result that $t_c$ diverges in the thermodynamic limit is not changed by the finite particles' size effects. The liquid phase influences the solution in the dilute region only by a minor correction to the effective length of the channel (the wall becomes located not at the end of the channel, but at the beginning of the liquid phase), hence the uniform decay rate of the pressure which value can be found considering the dilute phase is approximately the same as in the dilute case.

The study assumes that the solid phase does not form in the system, so that the fluid mechanics holds. Though this seems reasonable due to the growth of the pressure when the density becomes comparable with $\sigma^{-3}$, cf. \cite{Fouxon1,Fouxon2,Puglisi}, this question has to be studied. It is left for the future work. 

We conclude that for large system size the evolution reminds the gas-liquid transition. If one starts with the uniform initial state of the dilute gas, the formation of dense regions starts due to the clustering instability. The system develops the IS where the liquid condenses in a small part of the total container that takes almost all the mass of the system. The larger fraction of the system's volume is occupied by the dilute gas. At the boundary between the two phases the no-heat flux b. c. holds approximately so that the gas state is the same as would hold if the liquid would be the wall. Though the mass of the dilute gas is only logarithmic in the system size, the gas phase carries most of the energy of the system via the high velocity of its particles. 

\label{dense}

\section{Finite-time singularity regularized by the IS}
\label{finite} 

In the recent work \cite{Kolvin} the numerical simulations of the fluid-mechanical equations (\ref{fl}) of the dilute granular gas was performed in two dimensions. The results indicate the possibility of the finite-time singularity. This is quite plausible physically since the heat conduction coefficient that counterbalances the non-linear growth of the density due to cooling would tend to zero at such a presumed singularity. Indeed, if the pressure remains finite at the singularity, which seems to be the case, then the temperature tends to zero inversely proportionally to the growth of the density. Though in the one-dimensional case the heat conduction does stop the growth of the density, in the higher-dimensional case, where there are wider geometric possibilities for the formation of regions of growing density, this might be not the case. In fact, this is indicated by the analogy between the IS solutions and the soliton solutions of the non-linear physics. Within the latter there are cases 
where in dimension higher than one, the non-linearity produces finite-time singularities that cannot be stopped by the Laplacian terms in the equations. 

Thus the conjecture that the density of the dilute granular gas becomes singular in finite-time when the container's geometry is a box (which is described by fluid-mechanics of dimension higher than one) is reasonable.
This increases further the relevance of our derivation of the IS in the dense case. The finite-time singularity signifies that the frame of the dilute granular gas is inconsistent in dimension higher than one, so that physical factors not included into that frame have to be taken into account. The immediate factor is the finite size of the particles and the related excluded volume effects. It is clear that the fluid-mechanics of (possibly) dense fluid of hard spheres, described by Eqs.~(\ref{constf}),(\ref{fluideq}), does not have finite-time singularities becoming then ${\it the}$ only consistent framework of consideration in the higher-dimensional case. The IS solutions (including the uniform dense solution) become then highly important as the reference solutions on which further theoretical and experimental study can rely.

\section{Conclusion}

We described the IS states of the dense fluids of inelastically colliding hard-core particles. Though we used the fluid mechanics, the IS are not really fluid mechanical: they involve no flow. The inhomogeneity of the temperature is preserved by the balance of heat conduction and inhomogeneous inelastic cooling.  The IS are exact solutions: they solve the complete system of the coupled PDE of the fluid mechanics of the system. Though the precise form of the coefficients of those equations is unknown in the dense region, we succeeded to demonstrate the IS using only the special separable form of those coefficients that holds for hard spheres with constant coefficient of normal restitution.

The inelastic cooling obeys the power-law $[1+t/t_c]^{-2}$ where $t_c$ becomes infinite in the thermodynamic limit. The growth of the cooling time (that by itself is determined by the local density and temperature) with the system's size signifies that the whole system is strongly correlated. The existence of the non-trivial steady state in the dissipative system (the trivial one being the frozen particles) is unusual. It poses for the study the question whether self-organization and the minimization of the dissipation can be related in the considered case. 

The IS have universal properties that hold independently of the constitutive relations of the coefficients of the fluid-mechanical equations. The pressure and the energy decay as $[1+t/t_c]^{-2}$ while the number of collisions that occurred in the system and minus the entropy increase as $\ln(1+t/t_c)$. The only unknown characteristic of the IS is the form of the density field that does depend on the form of the 
coefficients. Thus the IS exhibit many universal properties that make one suggest that the displayed physical mechanisms can be important in other situations. 

The IS solutions depend on the absence of the characteristic energy scale in the problem: the interaction of the hard spheres involves no energy scale. The IS would not exist as exact solutions for finite interaction potential (instead of the infinite step potential of the hard spheres) that possesses a certain scale of energy, or for inelasticity which law changes at a certain scale of the energy. In the former case the coefficients of the fluid mechanics would have unknown dependence on the temperature, while in the latter case the coefficient of the inelastic energy loss term would have unknown dependence on the temperature. Nevertheless, it seems that the considered model can describe realistically certain regimes of evolution of the granular media, arising as intermediate asymptotic regime.

We introduced a transformation that transforms the IS into the time-independent solutions of a system of PDE that does not depend on time explicitly. Thus one can pass to the "frame of the IS" where the solutions are stationary. This is done by using the time variable which is the number of collisions that occurred in the system, and rescaling the fields with time to compensate for the decays due to inelasticity. In particular the transformation shows that the linear perturbations near the IS obey the power-law behavior in time. 

To consider the IS further we studied the dilute granular gas in the channel, where the fluid mechanical fields depend only on the spatial coordinate along the channel (the microscopic motion is still three dimensional for balls and two-dimensional for discs). While the IS in this case are known from the previous work \cite{MFV,IF}, their stability was known only in the case of not too long channels. In the limit of large size the density field of the IS has large variation where it changes from a small value (that vanishes in the thermodynamic limit of infinite length of the channel) to a value $\rho_{max}$ that is exponentially large in the channel's length. The interpolation between the two regions follows the inverse linear law. The mass contained in the neighborhood $\rho\sim\rho_{max}$ is of order one.  These solutions hold if the diluteness condition $\rho_{max} \sigma^3\ll 1$ holds. When the length of the channel is fixed, this condition holds if the particles' diameter $\sigma$ is small.  

We showed numerically that the IS provide the universal long-time limit of the evolution of the gas when the length of the channel exceeds the critical length $l_{cr}$. To consider the thermodynamic limit for finite-size particles, where dense liquid regions appear in the fluid, we demonstrated the phase separation in IS. In the limit of large system size, the fluid separates into the liquid phase, that contains most of the mass of system, and the gaseous phase that contains most of the energy of the system and occupies the volume's fraction close to unity. Since there is local stability in both phases, then it follows that the IS is globally stable and constitutes the result of the long-time evolution of the system. Thus for the first time the question of the long-time limit of the granular gas is settled completely, though for a special geometry of the container.

Our study shows that the IS play crucial role in the behavior of the granular fluid of hard spheres in the channel. The IS however hold in any geometry of the container. 
Their relevance to the case where the fluid mechanical fields depend on two or three spatial variables is important subject of future work. The numerical works reported in \cite{Kolvin} indicate that the 
dilute granular gas develops infinite density in finite time if the density depends on two coordinates. The IS that take into account the excluded volume effects, do not have such singularities and become 
important objects for the study of the evolution in the box geometry of the container. This study is left for future work. 

We would like to thank J. Vollmer and A. Vilenkin for discussions and help without which this work would not appear. 

This work was supported by the grant of the Minerva Foundation with funding from the German Ministry for Education and Research at the Weizmann Institute of Science

\end{document}